\documentclass[aps,showpacs,superscriptaddress,nofootinbib,amsmath,amssymb]{revtex4}
 \usepackage{graphicx}
\usepackage{dcolumn}
\usepackage{bm}
\usepackage{lipsum}
\usepackage{adjustbox}
\usepackage{multirow}
\graphicspath{./fig}
\begin{document}
 \title{Nucleon and nuclear structure functions with non-perturbative and higher order perturbative QCD effects}
\author{F. Zaidi}
\affiliation{Department of Physics, Aligarh Muslim University, Aligarh - 202002, India}
\author{H. Haider\footnote{on leave from Aligarh Muslim University}}
\affiliation{Fermi National Accelerator Laboratory, Batavia, Illinois
60510, USA}
\author{M. Sajjad Athar\footnote{Corresponding author: sajathar@gmail.com}}
\author{S. K. Singh}
\affiliation{Department of Physics, Aligarh Muslim University, Aligarh - 202002, India}
\author{I. \surname{Ruiz Simo}}
\affiliation{Departamento de F\'{\i}sica At\'omica, Molecular y Nuclear,
and Instituto de F\'{\i}sica Te\'orica y Computacional Carlos I,
Universidad de Granada, Granada 18071, Spain}
\begin{abstract}
We have studied the nucleon structure functions $F_{iN}^{EM} (x,Q^2);~i=1,2$, by including contributions due to the higher order perturbative QCD effect up to NNLO 
 and the
 non-perturbative effects due to the kinematical and dynamical higher twist (HT) effects. The numerical results for $F_{iN}^{EM}(x,Q^2)$ are obtained using Martin, 
 Motylinski, Harland-Lang, Thorne (MMHT) 2014 NLO and NNLO nucleon parton distribution functions (PDFs). The dynamical
  HT correction has been included following the renormalon approach as well as the phenomenological approach and the kinematical HT effect is incorporated using
  the works of Schienbein et al. These nucleon
  structure functions have been used as an input to calculate the nuclear structure functions $F_{iA}^{EM} (x,Q^2)$. 
   In a nucleus, the nuclear corrections arise because of the Fermi motion, binding energy, 
   nucleon correlations, 
 mesonic contribution, shadowing and antishadowing effects. These nuclear corrections are taken into account in the numerical calculations
 to obtain the nuclear structure functions $F_{iA}^{EM} (x,Q^2)$, for the various nuclear targets like 
 $^{12}C$, $^{27}Al$, $^{56}Fe$, $^{64}Cu$, $^{118}Sn$, $^{197}Au$ and $^{208}Pb$ which are of experimental interest.  
 The effect of isoscalarity correction for nonisoscalar nuclear targets has also been studied.
 The results for the $F_{iA}^{EM} (x,Q^2)$ are compared with nCTEQ nuclear PDFs parameterization as well as 
 with the experimental results from JLab, SLAC and NMC in the kinematic region of $0.1 \le x \le 0.8$ for several nuclei. 
  We have also calculated the ratio $R_{A} (x,Q^2)
 =\frac{F_{2A}(x,Q^2)}{2 x F_{1A}(x,Q^2)}$ in the moderate $Q^2$ region for various nuclei and compared the results with the 
 available experimental data from JLab to examine
 the validity of Callan-Gross relation in the nuclei. We also make predictions for the nuclear structure functions 
 in $^{12}C$, $^{64}Cu$ and $^{197}Au$ in the kinematic region of the proposed experiment at the JLab.
  \end{abstract}
\maketitle
 \section{Introduction} 
 A better theoretical understanding of the nuclear medium effects in the deep inelastic scattering(DIS) region 
  in the electromagnetic(EM) and weak interaction induced processes has been emphasized
  ~\cite{Alvarez-Ruso:2017oui, Bodek:2015mwa, Guzey:2015yk, Solvignon:2009it, Kovarik:2012zz, Haider:2016zrk, Haider:2015vea}
  in view of the present experiments being performed on various nuclear targets
 using electron beam at the JLab~\cite{Mamyan:2015th, jlab, jlab_current, Covrig} and the neutrino/antineutrino beams at the Fermi lab~\cite{Fermilab}. 
 A dedicated experiment at the JLab 
 to study the nuclear medium effects in the kinematic
 region of $1<Q^2<5~GeV^2$ and $0.1<x<0.6$ for the electron induced DIS process on $^1H$, $^2D$, $^{12}C$, $^{64}Cu$ and $^{197}Au$
 targets has been proposed~\cite{Covrig}. In the scattering of charged lepton from nucleon target, the region of high energy(or large $Q^2$) in the DIS
 is well described by the perturbative
 Quantum Chromodynamics(pQCD).
 However, in the few GeV energy range, or equivalently moderate $Q^2$,
 where the hadronic degrees of freedom are dominant, the strong coupling constant becomes large and the application of perturbative QCD becomes inadequate. 
 This is the energy region 
 where it is easier to work with the hadronic degrees of freedom using resonances.
  The region of moderate $Q^2$ is also known as the shallow inelastic region (SIS) or the transition region (the region between
 the $\Delta$-resonance production and the DIS region i.e. $W > 2~GeV,~Q^2 > 1~GeV^2$). 
  In the transition region besides $\Delta(1232)$ resonance, there are several higher resonances like 
 P$_{11}$(1440), D$_{13}$(1520), S$_{11}$(1535), S$_{11}$(1650), P$_{13}$(1720), etc. which contribute to the event rates. In this region, except for a few resonances,
 limited informations are available on the transition form factors and coupling strengths, etc. which are needed to calculate 
 the contribution of these resonances to the event rates.
 The study of the shallow inelastic region, is important to understand the hadronic interactions, 
 for the electromagnetic as well as the weak processes. Attempts are made to understand this shallow inelastic region 
 in terms of quark-hadron duality which describes a remarkable similarity between the electron-nucleon scattering in the DIS region, 
 where the electron scattering from an asymptotically free point like quark is assumed to take place, and the nucleon
 resonance region where the electron nucleon scattering takes place with a correlated cluster of quarks and gluons. 
 The phenomenon of the quark-hadron duality was first observed by Bloom and Gilman~\cite{Bloom:1970xb} while analyzing SLAC data, which 
 showed a striking similarity between the $F_2(x,Q^2)$ structure function measured in the resonance region and the DIS region.
 The phenomenon of quark-hadron duality, 
 therefore, may play an important role in the understanding of electron-nucleon scattering in this region.
 When electron scattering
 takes place with a bound nucleon in a nuclear target like $^{12}C$, $^{64}Cu$, $^{197}Au$, etc., nuclear medium effects (NME) become important which was first observed 
 by the EMC experiment and later confirmed by other experiments showing that the nucleon structure functions 
 $F_{iN}^{EM} (x,Q^2);~i=1,2$, is considerably modified in the nuclear medium which is of interest to the nuclear physics community. 
 
 The study of SIS region is also important in the
 neutrino/antineutrino experiments being performed in the few GeV energy region. Almost all the  neutrino/antineutrino experiments are using moderate to heavy 
 nuclear targets like $^{12}C$, $^{16}O$, $^{40}Ar$, $^{56}Fe$ and $^{208}Pb$.
 There is a dedicated experiment presently running at the Fermi lab (MINERvA)~\cite{Fermilab}, 
 where the nuclear medium effects are being studied using several nuclear targets in the
 $\nu_l/\bar\nu_l-$nucleus scattering, as well as there is plan to study NME in $^{40}Ar$ in the proposed DUNE experiment at 
 the Fermi lab~\cite{Mousseau:2016snl, Acciarri:2015uup}. These neutrino experiments are being performed in the few GeV energy region, where
 considerable uncertainty in the neutrino/antineutrino-nucleus cross sections ($\approx 25\%$) adds to the total systematics. 
 For example, the DUNE at the Fermi lab, is expected
 to have more than $50\%$ interactions by $\nu$ and $\bar\nu$ on the bound nucleons inside the nuclear targets, 
 in the transition region of the shallow inelastic(SIS) to DIS with $W$ above 
 the mass of the $\Delta$ resonance region~\cite{Acciarri:2015uup}. The importance of studying electron and neutrino/antineutrino interactions for nucleons and nuclear targets in the transition region has been
 emphasized recently in the conferences and workshops in the context of modeling $\nu(\bar\nu)$-nucleus interactions to analyze the ongoing 
 neutrino oscillation experiments~\cite{nustec}. Presently some phenomenological approach to extrapolate the DIS cross sections to lower energy region is used in 
 most of the neutrino event generators to obtain the neutrino/antineutrino-nucleus cross section in the transition region. 
 A good understanding of the SIS region in the electromagnetic scattering is essential in order to calculate the weak 
 cross sections induced by $\nu_l$ and $\bar\nu_l$ in this region.
 Therefore, in this paper, we have studied nuclear medium effects in the structure functions 
 at moderate $Q^2$ corresponding to the JLab kinematics in the SIS region. This study will be helpful in future attempts to apply this formalism in the 
 transition region to the
 weak interaction induced processes.
 
 Generally, the experimental results of the cross section for DIS processes induced by the
 charged leptons and the neutrino/antineutrino on the nucleons and the nuclear targets are interpreted in terms of the  
structure functions.  
In the case of EM DIS processes induced by the leptons on the nucleons, the cross section is expressed as
\begin{small}
\begin{equation}\label{eq1}
 \frac{d^2\sigma}{dxdy}=\frac{8\pi\alpha^2 M_N E}{Q^4}\left(xF_{1N} (x,Q^2) y^2 + F_{2N} (x,Q^2) \left[1-y-\frac{M_N x y}{2 E} \right] \right),
\end{equation}
\end{small}
 where $F_{1N} (x,Q^2)$ and $F_{2N} (x,Q^2)$ are the two nucleon structure functions, $x(=\frac{Q^2}{2 M_N \nu})$ is the Bjorken scaling variable,
 $y=\frac{\nu}{E}$, $M_N$ is the mass of target nucleon, $\nu(=E-E^\prime)$ and $Q^2(=4EE^\prime sin^2\left({\theta \over 2}\right))$ are the energy transfer and four momentum transfer square to the hadronic system and $E(E')$ is the incident(outgoing) energy of the lepton. 
The structure function $F_{1N}(x,Q^2)$ describes the contribution of the transverse component of the virtual photon to the DIS cross sections
while the structure function $F_{2N}(x,Q^2)$ describes a linear combination of the longitudinal and transverse
components. Alternately, the DIS cross section is also described in terms of the transverse structure function $F_{TN}(x,Q^2)$ 
and the longitudinal structure function $F_{LN}(x,Q^2)$ defined as
\begin{equation}
F_{TN} (x,Q^2)= 2x F_{1N}(x,Q^2)\;;\;\;\;\;
 F_{LN} (x,Q^2)=\left(1+\frac{4 M_N^2 x^2}{Q^2} \right) F_{2N} (x,Q^2) - 2 x  F_{1N} (x,Q^2).
\end{equation}
The transverse and longitudinal cross sections are then expressed as
\begin{equation}
     \sigma(x,Q^2) = \sigma_{TN}(x,Q^2) +  \sigma_{LN}(x,Q^2), 
\end{equation}
where
\begin{equation}
     \sigma_{TN,LN}(x,Q^2) = \left(\frac{4\pi^2\alpha}{2x\nu(1-x)M_N} \right) F_{TN,LN} (x,Q^2). 
\end{equation}

 The ratio of nucleon structure functions, $R_N(x,Q^2)$ is defined as
\begin{eqnarray}\label{rlnr2n}
 R_{N} (x,Q^2)= {F_{2N} (x,Q^2) \over 2 x F_{1N} (x,Q^2)}.
\end{eqnarray}
  In the kinematic region of Bjorken scaling($Q^2 \to \infty,~\nu \to \infty$ such that $x={Q^2 \over 2M_N\nu} \to$constant), all
the nucleon structure functions scale i.e. $F_{iN} (x,Q^2) \rightarrow F_{iN} (x)\;\;(i=1,2,L) $. In this kinematic region, the
structure functions $F_{1N} (x) $ and $F_{2N} (x)$ calculated in the quark-parton model satisfy the Callan-Gross relation(CGR) given by~\cite{Callan:1969uq}:
\begin{eqnarray}
 \label{cgr}
   F_{2N} (x) &=& 2 x F_{1N} (x)\\
   \label{rr}
\mbox{implying}\;\;\; R_{N}(x,Q^2) &\rightarrow& 1,\;\mbox{in the limit of $Q^2$} \rightarrow \infty.
\end{eqnarray}

Therefore, in the kinematic limit of the Bjorken scaling, the EM DIS data on the scattering of the electrons from the proton targets are analyzed in terms of 
 only one structure function $F_{2N} (x)$. An explicit evaluation of $F_{2N} (x)$ in the quark parton model gives~\cite{cooper_sarkar}: 
 \begin{equation}
\label{eq:f22xf1}
F_{2N} (x) = 2 x F_{1N} (x) = x \sum_i e_i^2 \left( f_i(x) + \bar f_i(x) \right),
\end{equation}
where $f_i(x)$ and $\bar f_i(x)$ are the quark and antiquark parton distribution functions(PDFs) which describe the probability of finding a quark/antiquark of
flavor $i$ carrying a momentum fraction $x$ of the nucleon's momentum. $e_i$ is the charge corresponding to the quark/antiquark of flavor $i$.

 As we move away from the kinematic region of the validity of Bjorken scaling 
 towards the region of smaller $Q^2$ and $\nu$, the description 
 of the structure functions becomes more difficult to understand as there are various effects that come into play like the target mass 
  correction(TMC) and the higher twists(HT),
 as well as other non perturbative
QCD effects arising due to the quark-quark and quark-gluon interactions which are expected to give rise to $Q^2$ dependent contribution to the structure 
 functions. This results in the violation of Bjorken scaling.
 Theoretical studies show that the corrections to the 
 nucleon structure functions due to these effects decrease as ${1 \over Q^2}$, and therefore become important at small and moderate
 $Q^2$~\cite{Melnitchouk:2005zr, Schienbein:2007gr,Miramontes:1989ni, Castorina:1984wd, Melnitchouk:2006je}. 
 These contributions may be different for $F_{1N}(x,Q^2)$ and $F_{2N}(x,Q^2)$ leading to different $Q^2$ dependent corrections in CGR given by Eqs.(\ref{cgr}) 
 and (\ref{rr}). 
There exist some phenomenological attempts to study the deviation of $F_{LN}(x,Q^2) \over {2 x F_{1N}(x,Q^2)}$ from its Bjorken limit 
 by studying the $Q^2$ dependence of $F_{LN}(x,Q^2)$ in the region of smaller and moderate $Q^2$~\cite{Whitlow:1991uw, Bodek:2010km, Christy:2007ve, Bosted:2015qc, 
 Abe:1998ym, Dasu:1988ms}.
 These phenomenological studies describe the available experimental results on 
 $F_{LN}(x,Q^2) \over {2 x F_{1N}(x,Q^2)}$~\cite{Whitlow:1991uw, Bodek:2010km, Christy:2007ve, Abe:1998ym, Dasu:1988ms, Tvaskis:2006tv, Monaghan:2015et, Liang:2004tj,
 Benvenuti:1989rh}.
 The most widely used parameterization of this ratio $\left(F_{LN}(x,Q^2) \over {2 x F_{1N}(x,Q^2)}\right)$ is given by Whitlow et al.~\cite{Whitlow:1991uw}.

In the case of nuclear targets, the EM DIS cross sections are similarly analyzed in terms of the nuclear structure function $F_{2A} (x,Q^2)$
assuming the validity of CGR at the nuclear level.
A comparative study of the nuclear structure function $F_{2A}(x,Q^2)$ with the free nucleon structure function $F_{2N}(x,Q^2)$ 
 led to the discovery of the EMC effect~\cite{Aubert:1983xm, Bodek:1983qn, Bodek:1983qn}. The nuclear medium effects arising due to the  
 Fermi motion, binding energy, nucleon
 correlations, shadowing, etc., in understanding the EMC effect, in the various regions of $x$ has been extensively studied
 in the last 35 years~\cite{Malace:2014uea, Geesaman:1995yd, Hen:2013oha}.
 
However, there have been very few theoretical attempts to make a comparative study of the nuclear medium effects in $F_{1A}(x,Q^2)$, $F_{2A}(x,Q^2)$ and 
 $F_{LA}(x,Q^2)$,
 and understand their modifications in nuclei. The recent experimental measurements on the EM nuclear structure functions reported from the JLab on various nuclei in the kinematic region of
 $Q^2(1<Q^2<5~GeV^2)$ and $x(0.1<x<1)$ also show that the nuclear medium effects are different 
 for  $F_{1A}  (x,Q^2)$,  $F_{2A}  (x,Q^2)$ and  $F_{LA}  (x,Q^2)$, which could modify the CGR in nuclei~\cite{Mamyan:2015th}. 

 In view of these experimental results a theoretical study of the nuclear structure functions $F_{iA} (x,Q^2)(i=1,2,L)$ for the electromagnetic processes 
 and its effect on $R_{A}(x,Q^2)=\frac{F_{2A}(x,Q^2)}{2 x F_{1A}(x,Q^2)}$ and CGR in the nuclear medium in the various regions of $x$ and $Q^2$ is highly desirable. 
 A comparison of the theoretical results 
 with the present and future experimental data
 from the JLab~\cite{Mamyan:2015th,jlab,jlab_current,Covrig} will lead to a better 
 understanding of the nuclear medium effects in the EM structure functions.\\

 In this work, we have studied the following aspects of the structure functions:
 \begin{itemize}
 \item  The free nucleon structure functions $F_{iN} (x,Q^2)$ ($i=1,2,L$) have been numerically calculated using the nucleon PDFs of
 Martin, Motylinski, Harland-Lang, Thorne (MMHT)~\cite{Harland-Lang:2014zoa}. For the evolution of PDFs at the next-to-leading order(NLO) and next-to-next-to-leading order(NNLO) obtained from the leading order(LO),
 we have followed the works of Vermaseren et al.~\cite{Vermaseren:2005qc} and Moch et al.~\cite{Moch:2004xu} and obtain the nucleon structure functions
 $F_{1N}(x,Q^2)$ and $F_{2N}(x,Q^2)$ independently. 
 The target mass correction effect has been included
 following the method of Schienbein et al.~\cite{Schienbein:2007gr}. 
 The dynamical higher twist correction has been taken into account following the methods of Dasgupta et al.~\cite{Dasgupta:1996hh} and 
 Stein et al.~\cite{Stein:1998wr} as well as the phenomenological approach of Virchaux et al.~\cite{Virchaux:1991jc}. 
 
  \item The nuclear medium effects arising due to the Fermi motion, binding energy, nucleon correlations
 have been taken into account through the use of spectral function of the nucleon in the nuclear medium. In
 addition to that we have incorporated mesonic contributions due to $\pi$ and $\rho$ mesons~\cite{Marco:1995vb,GarciaRecio:1994cn}, shadowing and antishadowing effects~\cite{Kulagin:2004ie}. For the pionic PDFs we have used the parameterizations given by Gluck et al.~\cite{Gluck:1991ey}
 and also made a comparative study by using the pionic PDFs parameterization given by Wijesooriya et al.~\cite{Wijesooriya:2005ir}. For the 
 rho mesons the same PDFs as for the pions have been considered.
 
 \item The nuclear corrections in the structure functions $F_{iA} (x,Q^2)$ ($i=1,2,L$) and the
 nuclear dependence on
 $R_{A}(x,Q^2)=\frac{F_{2A}(x,Q^2)}{2 x F_{1A}(x,Q^2)}$, have been studied  
 in the regions of $Q^2$ and $x$ relevant for the experiments which have been performed in the nuclei like $^{12}C$, $^{27}Al$, $^{56}Fe$, $^{64}Cu$ etc.
 \cite{Mamyan:2015th}. 
 The results are compared with the available experimental data from the JLab~\cite{Mamyan:2015th}, 
 SLAC~\cite{Gomez:1993ri} and NMC~\cite{Arneodo:1996rv} experiments.   
 The results are also compared with those obtained with the phenomenological nCTEQ nuclear PDFs parameterization~\cite{Kovarik:2015cma}.
 The predictions have been made in the kinematic region relevant to the 
 future experiments to be performed at the JLab, in several nuclei like $^{64}Cu$, $^{197}Au$, etc.~\cite{Mamyan:2015th}.
 
 \item The results for the nonisoscalar($N>>Z$) nuclear targets are compared with the results when these nuclei are treated as isoscalar targets, to study
 the effect of isoscalarity correction.
 We have also studied the $W$ dependence (where $W$ is the center of mass energy of the final hadronic state), of nuclear structure functions.
 This is important to understand the 
 $x$ and $Q^2$ dependence of the structure functions in the transition region from resonance to DIS.
 \end{itemize}
 
 In section \ref{sec_formalism}, the formalism for calculating the electromagnetic
structure functions and the ratio $R_A(x,Q^2)$ in the nuclear 
medium is given in brief. In section \ref{sec_results}, 
 the numerical results are presented.
 
\section{Formalism}\label{sec_formalism}
In a nucleus, the charged lepton interacts with the nucleons which are moving
with some momenta constrained by the Fermi momentum and Pauli blocking, 
 and the nucleus is assumed to be at rest. Therefore, the free
 nucleon quark and antiquark PDFs should be convoluted with the momentum distribution of the nucleons.
 In addition, there are binding
energy corrections. Furthermore, the target nucleon being strongly interacting particle interacts with the other   
 nucleons in the nucleus leading to the nucleon correlations. We have taken these effects into account by using a field theoretical model which starts 
 with the Lehmann's representation for the relativistic nucleon propagator and the nuclear many body theory is used to calculate it for
 an interacting Fermi sea in the nuclear matter. A local density approximation is then applied to obtain the results for 
 a finite nucleus. This technique results into the use of a relativistic nucleon 
spectral function that describes the energy and momentum distributions~\cite{FernandezdeCordoba:1991wf}. 
 All the information like Fermi motion, binding energy and the nucleon correlations is contained in the spectral function. 
 Moreover, we have considered the contributions of the pion and rho mesons 
in a many body field theoretical approach 
based on Refs.~\cite{Marco:1995vb,GarciaRecio:1994cn}. 
 The free meson propagator is replaced by a dressed one as these mesons interact with the nucleons in the
nucleus through the strong interaction.
We have earlier applied this model to study the nuclear medium effects
in the electromagnetic and weak processes~\cite{Haider:2016zrk, Haider:2015vea, Haider:2014iia, Haider:2011qs, Haider:2015nf, Haider:2015ic, SajjadAthar:2009cr}, 
as well as proton induced Drell-Yan processes~\cite{Haider:2016tev} on the nuclear targets.
\subsection{Lepton-Nucleon Scattering}\label{lpn}
For the charged lepton induced deep inelastic scattering process 
($l(k) + N(p) \rightarrow l(k^\prime) + X(p^\prime);$ $l=~e^-,~\mu^-$), 
the differential scattering cross section is given by
\begin{equation}\label{eAf}
\frac{d^2 \sigma_N}{d\Omega dE^{\prime}} =~\frac{\alpha^2}{q^4} \; \frac{|\bf k'|}{|\bf k|} \;L_{\mu \nu} \; W_N^{\mu \nu},
\end{equation}
where $L_{\mu \nu}$ is the leptonic tensor and the hadronic tensor $W_N^{\mu \nu}$ is defined in terms of 
nucleon structure functions $W_{i N}$(i=1,2) as
\begin{eqnarray}\label{nuclearhtf}
L_{\mu \nu} &=& 2 [k_{\mu} k'_{\nu} +  k'_{\mu} k_{\nu} - k\cdot k^\prime g_{\mu \nu}] \nonumber\\
W_N^{\mu \nu} &=& 
\left( \frac{q^{\mu} q^{\nu}}{q^2} - g^{\mu \nu} \right) \;
W_{1N} + \left( p_N^{\mu} - \frac{p_N . q}{q^2} \; q^{\mu} \right)
\left( p_N^{\nu} - \frac{p_N . q}{q^2} \; q^{\nu} \right)
\frac{W_{2N}}{M_N^2}
\end{eqnarray}
with $M_N$ as the mass of nucleon. 

In terms of the Bjorken variable $x\left(=\frac{Q^2}{2M_N\nu}=\frac{Q^2}{2M_N (E - E^{\prime})}\right)$ and
$y\left(=\frac{\nu}{E}\right)$,
where $Q^2=-q^2$ and $\nu$ is the
energy transfer($=E-E^\prime$) to the nucleon in the Lab frame $ \left(\nu=\frac{p_N\cdot q}{M_N}=\frac{p^{ 0}_N q^0-p_{N}^z q^z}{M_N}\right)$, 
the differential cross section is given by
\begin{eqnarray}\label{diff_dxdy1}
\frac{d^2 \sigma_N}{d x d y}&=&
\frac{\pi \alpha^2 M_N^2 y}{2 E E^{\prime} sin^4\frac{\theta}{2}}
\left\{W_{2 N}(x, Q^2)cos^2\frac{\theta}{2}~+~2W_{1 N}(x, Q^2)sin^2\frac{\theta}{2}   \right\}\,.
\end{eqnarray}

Expressing in terms of dimensionless structure functions $F_{1N}(x,Q^2)=M_N W_{1 N}(\nu,Q^2)$ and $F_{2 N}(x,Q^2)=\nu W_{2 N}(\nu,Q^2)$, this is equivalent to Eq.\ref{eq1}.
%

The partons inside the nucleon may interact among themselves via gluon exchange which is described by the QCD. For example, through the channels 
$\gamma^\ast g \to q \bar q$ and $\gamma^\ast q \to q g$, if one takes into account the contribution from the gluon emission then the nucleon structure function shows dependence on $Q^2$, i.e. 
 Bjorken scaling is violated. The $Q^2$ evolution of
structure functions is made by using the Altarelli-Parisi evolution equation~\cite{Altarelli:1977zs}.
In the limit of $Q^2 \to \infty$, the strong coupling constant $\alpha_s(Q^2)$ becomes very small and therefore, the higher order terms can be neglected in
comparison to the leading order term. 
But for a finite value of $Q^2$, $\alpha_s(Q^2)$ is large and next-to-leading order terms give a significant contribution followed by next-to-next-to-leading 
order terms. The details of method to incorporate evolution are given in
Refs.~\cite{Vermaseren:2005qc, Neerven, Furmanski:1981cw, Hirai:1997gb, Kumano:1992vd, Coriano:1998wj, Ratcliffe:2000kp, Moch:2004xu}. 

In this work, we have used the MMHT 2014 PDFs for the nucleons at NLO and NNLO~\cite{Harland-Lang:2014zoa}.
The nucleon structure functions $F_2(x,Q^2)$ and $F_L(x,Q^2)$ are expressed as~\cite{Neerven,Moch:2004xu}:
\begin{equation}\label{F_L}
x^{-1}F_{2,L}(x)=\sum_{f=q,g} C_{2,L}^{(n)}(x) \otimes f(x), 
\end{equation}
where $C_{2,L}$ are the coefficient functions for the quarks and gluons~\cite{Neerven,Moch:2004xu}, the superscript $n=0,1,2,3....$ for N$^{n}$LO, the symbol $\otimes$ represents the Mellin convolution and $f$ represents the quark and gluon distributions~\cite{Harland-Lang:2014zoa}.

At low $Q^2$, say a few $GeV^2$, in addition to higher-order QCD corrections~\cite{Alekhin:2007fh}, non-perturbative phenomena become important. 
 In the formalism of the operator product expansion (OPE), structure 
functions are generally expressed in terms of powers of $1/Q^2$ (power corrections), i.e.,  
\begin{equation}
F_{i}(x,Q^2) = F_{i}^{\tau = 2}(x,Q^2)
+ {H_{i}^{\tau = 4}(x) \over Q^2} 
+ {H_{i}^{\tau = 6}(x) \over Q^4} + .....   \;\;\; i=1,2,
\label{eqn:ht}
\end{equation}
 where the first term ($\tau=2$) is known as the leading twist (LT) term, and is responsible for the evolution 
of structure functions via perturbative QCD $\alpha_s(Q^2)$ corrections. 
The higher twist (HT) terms with $\tau = 4,6$,\ldots
reflect the strength of multi-parton correlations ($qq$ and $qg$). 
 Due to their nonperturbative origin, current models can only provide a 
qualitative description for such contributions, which is usually determined via
reasonable assumptions from data~\cite{Alekhin:2013nda,Accardi:2016qay}. In literature, various parameterizations are available for the HT contribution~\cite{Virchaux:1991jc, Martin}.

If the structure functions are evaluated at NNLO, then 
 most of the higher twist contributions extracted in the NLO fit at low $Q^2$ appear to simulate from the missing NNLO terms, 
 i.e. the magnitude of higher twist terms decreases
strongly when going from LO to NLO, and then to NNLO approximations to the evolution equation~\cite{Yang,Martin,KulaginS}. Moreover, an
 additional suppression of higher twist terms occurs when the nuclear effects are applied~\cite{KulaginS}. 

In addition to the dynamical HT terms defined in Eq.\ref{eqn:ht}, there are also kinematic HT contributions associated with
the finite mass of the target nucleon $M_N$, which are relevant at high $x$ and moderate $Q^2$. 
 The TMC arises due to the production of heavy quarks, like charm, bottom and top quarks through the photon-gluon, quark-gluon, gluon-gluon fusion etc., and their masses can not be ignored as compared to the nucleon mass. This results in the modification of the kinematics for the scattering process. 
  We have followed the  prescription of Schienbein et al.~\cite{Schienbein:2007gr}, where the Bjorken variable $x$ is replaced 
by the Nachtman variable $\xi$ defined as
\begin{eqnarray}
 \xi=\frac{2x}{1+\sqrt{1+\frac{4 M_N^2 x^2}{Q^2}}}
\end{eqnarray}
 and the expressions of structure functions including TMC effect are given by
\begin{eqnarray}\label{tmcf1f2}
 F_{2N}^{TMC}(x,Q^2) &\approx& {x^2 \over \xi^2 \gamma^3}~F_{2N}(\xi)~\left(1~+~6r(1-\xi)^2 \right)\nonumber\\
  F_{1N}^{TMC}(x,Q^2) &\approx& {x \over \xi r}~F_{1N}(\xi)~\left(1~+~2r(1-\xi)^2 \right),
\end{eqnarray}
where $r=\frac{\mu x \xi}{\gamma}$, $\mu=\frac{M_N^2}{Q^2}$ and $\gamma=\sqrt{1+\frac{4M_N^2 x^2}{Q^2}}$, respectively. 


\subsection{Lepton-Nucleus Scattering}\label{lpa}
 In the case of a nuclear target, the expression for the differential scattering cross section is given by
\begin{equation}\label{eA}
\frac{d^2 \sigma_A}{d\Omega dE^{\prime}} =~\frac{\alpha^2}{q^4} \; \frac{|\bf k'|}{|\bf k|} \;L_{\mu \nu} \; W_A^{\mu \nu},
\end{equation}
where $\alpha$ is the fine structure constant, $L_{\mu\nu}=2 \left( k_\mu k'_\nu + k'_\mu k_\nu - g_{\mu\nu} k \cdot k' \right)$ 
is the leptonic tensor and $W_A^{\mu \nu}$ is the nuclear hadronic tensor which is expressed in terms of the nuclear structure functions $W_{iA} (\nu,Q^2)(i=1,2)$ as
\begin{equation}
 W^{\mu\nu}_A
= W_{1A} (\nu,Q^2)
  \left( { q^\mu q^\nu \over q^2} - g^{\mu\nu} \right)
+ { W_{2A} (\nu ,Q^2) \over M_A^2 }
  \left( p_A^\mu - { p_A\cdot q \over q^2 } q^\mu \right)
  \left( p_A^\nu - { p_A\cdot q \over q^2 } q^\nu \right),
\label{eq:Wmunu_nucleus}
\end{equation}
where $M_A$ is the mass and $p_A$ is the four momentum of the target nucleus. 

The differential scattering cross section may also be written in terms of the 
probability per unit time ($\Gamma$) of finding a charged lepton interacting with a target nucleon given by~\cite{Haider:2016zrk, Haider:2015vea}:
\begin{equation}\label{defxsec1}
 d\sigma=\Gamma~dt~dS=\Gamma~\frac{dt}{dl}~dl~dS=\frac{\Gamma}{ v}dV=\Gamma\frac{E({\bf k})}{|{\bf k}|}dV=\frac{-2m_l}{\mid {\bf k} \mid} Im \Sigma (k) dV,
\end{equation}
\begin{figure}
 \includegraphics[height=3.5 cm, width=8.0 cm]{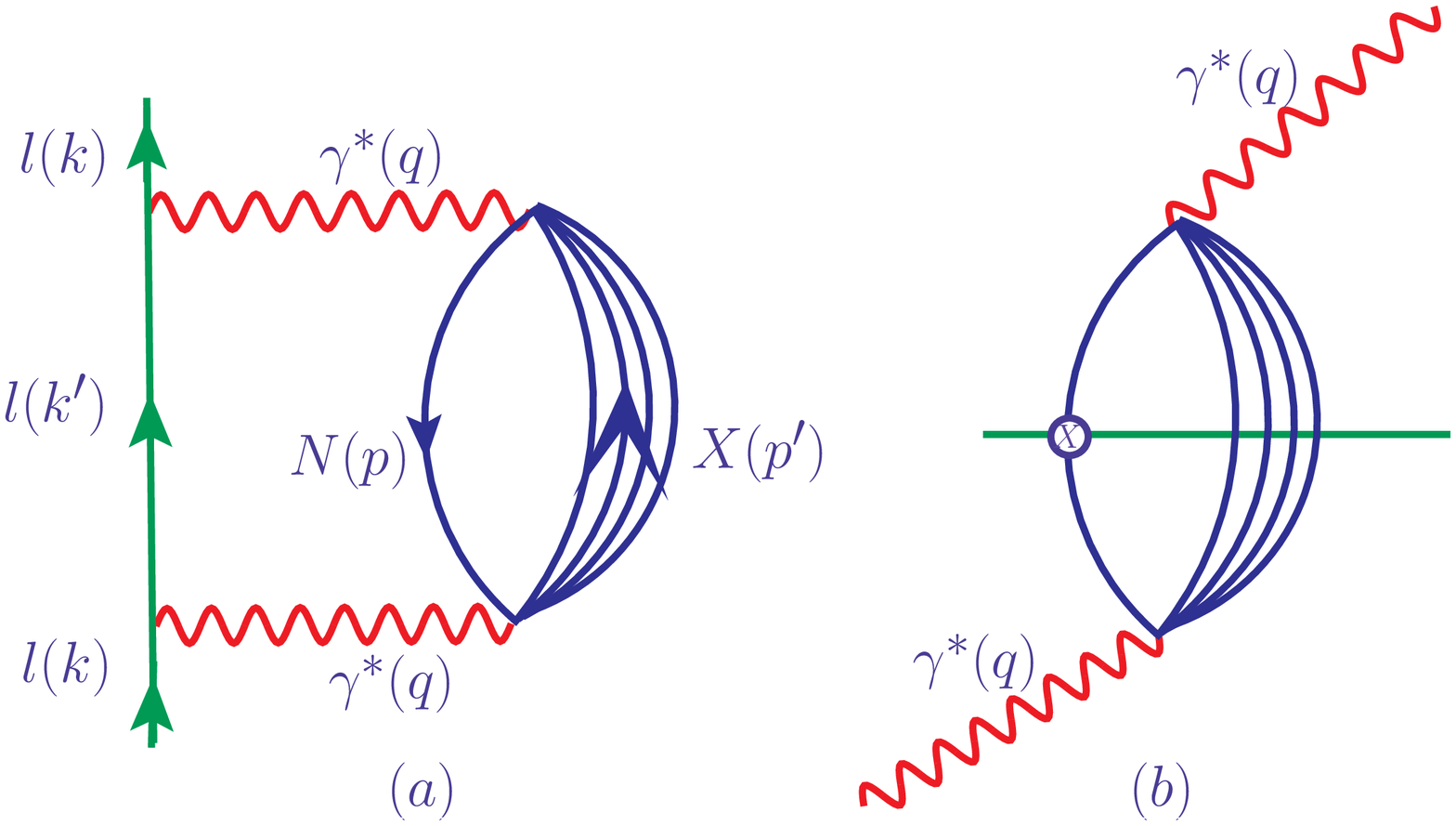}
 \caption{Diagrammatic representation of {\bf(a)} charged lepton self energy, {\bf (b)} photon self energy with Cutkosky cuts(solid horizontal line)
 for 
 putting particles on mass shell.}
 \label{lep-self}
\end{figure}
where $dt$ is the time of interaction, $dS$ is the differential area, $dl$ and $v(=\frac{|{\bf k}|}{E({\bf k})})$ stand for the length of interaction and velocity, 
respectively and $dV$ is the volume element inside the nucleus. $m_l$ is the lepton mass and
$Im \Sigma(k)$ is the imaginary part of the lepton self energy
(from the diagram of Fig.~\ref{lep-self}{\bf (a)})
which is obtained by using the Feynman rules for the lepton self energy($\Sigma(k)$) given by
\begin{equation}\label{defn}
\Sigma (k) = i e^2 \; \int \frac{d^4 q}{(2 \pi)^4} \;
\frac{1}{q^4} \;
\frac{1}{2m_l} \;
L_{\mu \nu} \; \frac{1}{k'^2 - m_l^2 + i \epsilon} \; \Pi^{\mu \nu} (q),
\end{equation}
where $\Pi^{\mu\nu}(q)$ is the photon self energy
which has been shown in Fig.~\ref{lep-self}{\bf (b)}.

Using Eq.(\ref{defn}) in Eq.(\ref{defxsec1}), the scattering cross section~\cite{Marco:1995vb} 
is obtained as
\begin{equation}\label{dsigma_3}
\frac {d^2\sigma _A}{d\Omega dE'}=-\frac{\alpha}{q^4}\frac{|\bf{k^\prime}|}{|\bf {k}|}\frac{1}{(2\pi)^2} L_{\mu\nu} \int  Im [\Pi^{\mu\nu}(q)] d^{3}r
\end{equation}
Now comparing Eq.(\ref{eA}) and Eq.(\ref{dsigma_3}), one may write the nuclear hadronic tensor $W_A^{\mu \nu}$ 
in terms of the photon self energy as:
\begin{eqnarray}\label{wamunu}
W_A^{\mu \nu}=-\frac{1}{4\pi^2\alpha} \int  Im [\Pi^{\mu\nu}(q)] d^{3}r
\end{eqnarray}

Using the Feynman rules, the
expression for $\Pi^{\mu\nu}(q)$ is obtained as
\begin{eqnarray}\label{photonse}
\Pi^{\mu \nu} (q)&=& e^2 \int \frac{d^4 p}{(2 \pi)^4} G (p) 
\sum_X \; \sum_{s_p, s_l} {\prod}_{\substack{i = 1}}^{^n} \int \frac{d^4 p'_i}{(2 \pi)^4} \; \prod_{_l} G_l (p'_l)\; \prod_{_j} \; D_j (p'_j)\nonumber \\  
&&  <X | J^{\mu} | H >  <X | J^{\nu} | H >^* (2 \pi)^4  \; \delta^4 (q + p - \sum^n_{i = 1} p'_i),\;\;\;
\end{eqnarray}
 where $G_l$ is the fermion propagator and $D_j$ is the boson propagator for particles in the final state denoted collectively by $X$. In the above expression, $<X | J^{\mu} | H >$ is the hadronic current; $s_p$ and $s_l$ are respectively, the spins of 
nucleon and fermions in the final hadronic state $X$. $G (p)$ is the relativistic nucleon propagator inside the nuclear medium which is 
obtained using perturbative expansion of Dyson series in terms of the nucleon self energy($\Sigma^N$) for an interacting Fermi sea. The nucleon self energy
may be obtained using many body field theoretical approach in terms of the spectral functions~\cite{FernandezdeCordoba:1991wf, Marco:1995vb}. Therefore, the
nucleon propagator $G (p)$ inside the nuclear medium may also be expressed in terms of the particle and hole spectral
functions as~\cite{FernandezdeCordoba:1991wf}:
\begin{eqnarray}\label{Gp}
G (p) =&& \frac{M_N}{E({\bf p})} 
\sum_r u_r ({\bf p}) \bar{u}_r({\bf p})
\left[\int^{\mu}_{- \infty} d \, \omega 
\frac{S_h (\omega, {\bf{p}})}{p_0 - \omega - i \eta}
+ \int^{\infty}_{\mu} d \, \omega 
\frac{S_p (\omega, {\bf{p}})}{p_0 - \omega + i \eta}\right]\,,
\end{eqnarray}
where $u$ and $\bar u$ are respectively the Dirac spinor and its adjoint, $\mu\left(=\frac{p_F^2}{2 M_N} + Re\left[ \Sigma^N{\tiny \left( \frac{p_F^2}{2 M_N}, p_F \right)} \right]\right)$ is the chemical potential and $p_F$
is the Fermi momentum. $S_h$ and $S_p$, respectively, stand for hole and particle spectral functions, the expression for which is taken from 
 Ref.~\cite{FernandezdeCordoba:1991wf}. The spectral functions contain the
information about the nucleon dynamics in the nuclear medium.
 All the parameters of the spectral function are determined by fitting the binding energy per nucleon and the Baryon number for each nucleus.
 Therefore, we are left with no free parameter. For more discussion please see Ref.~\cite{Haider:2016zrk, Haider:2015vea, Marco:1995vb}.

 To obtain the contribution to the nuclear hadronic tensor $W^{\mu \nu}_{A}$, which is coming from the bound nucleons i.e. $W^{\mu \nu}_{A,N}$, due to the 
 scattering of the charged leptons on the nuclear targets,
 we use Eq.(\ref{photonse}) and Eq.(\ref{Gp}) in Eq.(\ref{wamunu}), and express $W^{\mu \nu}_{A,N}$ 
 in terms of the nucleonic tensor $W^{\mu \nu}_{N}$ convoluted over the hole spectral function $S_h$, and get 
\begin{equation}	\label{conv_WA}
W^{\mu \nu}_{A,N} = 2 \sum_{\tau=p,n} \int \, d^3 r \, \int \frac{d^3 p}{(2 \pi)^3} \, 
\frac{M_N}{E ({\bf p})} \, \int^{\mu_\tau}_{- \infty} d p_0 S_h^\tau (p_0, {\bf p}, \rho^\tau(r))
W^{\mu \nu}_{\tau} (p, q), \,
\end{equation}
 where $\rho^\tau(r)$ is the proton/neutron density inside the nucleus 
  which is determined from the electron-nucleus scattering experiments
 and $S_h^\tau$ is the hole spectral function for the proton/neutron.

 We take the $zz$ component in Eq.(\ref{conv_WA}) for $W_{A,N}^{\mu\nu}$ and $W_{\tau}^{\mu\nu}$, 
the momentum transfer $\bf q$ along the $z$-axis, and using $F_{2N} (x) = \nu W_{2N} (\nu,Q^2)$, we obtain $ F_{{2A}} (x_A,Q^2)$ as~\cite{Haider:2016zrk, Haider:2015vea}:
 \begin{eqnarray} \label{em_f2_noniso}
F_{{2A,N}} (x_A,Q^2)  &=&  2\sum_{\tau=p,n} \int \, d^3 r \, \int \frac{d^3 p}{(2 \pi)^3} \, 
\frac{M_N}{E ({\bf p})} \, \int^{\mu_\tau}_{- \infty} d p_0 S_h^\tau (p_0, {\bf p}, \rho^\tau(r)) \times \left(\frac{M_N}{p_0~-~p_z~\gamma}\right) 
~F_{2\tau}(x_N,Q^2)\nonumber \\
&\times&\left[\frac{Q^2}{q_z^2}\left( \frac{|{\bf p}|^2~-~p_{z}^2}{2M_N^2}\right) +  \frac{(p_0~-~p_z~\gamma)^2}{M_N^2} \left(\frac{p_z~Q^2}{(p_0~-~p_z~\gamma) q_0 q_z}~+~1\right)^2\right], 
\end{eqnarray}
where $F_{2\tau}(x_N,Q^2);~(\tau=p,n)$ are the structure functions for the proton and neutron, calculated using quark-parton model.

 Similarly, taking the $xx$ component of the nucleon and nuclear hadronic tensors, and using $F_{1N} (x) = M_N W_{1N} (\nu,Q^2)$, we obtain $F_{{1A,N}} (x_A, Q^2)$ as
~\cite{Haider:2016zrk, Haider:2015vea}:
\begin{eqnarray}\label{conv_WA1}
F_{{1A,N}} (x_A, Q^2) &=& 2\sum_{\tau=p,n} AM_N \int \, d^3 r \, \int \frac{d^3 p}{(2 \pi)^3} \, 
\frac{M_N}{E ({\bf p})} \, \int^{\mu_\tau}_{- \infty} d p_0 S_h^\tau (p_0, {\bf p}, \rho^\tau(r)) \left[\frac{F_{1\tau}(x_N, Q^2)}{M_N}+ \frac{{p_x}^2}{M_N^2}
\frac{F_{2\tau}(x_N, Q^2)}{\nu_N}\right],~\;~~
\end{eqnarray}
where 
\begin{eqnarray}
x_N= \frac{Q^2}{2p \cdot q}= \frac{Q^2}{2(p_0q_0 - p_zq_z)}.
\end{eqnarray}
$F_{1\tau}(x_N,Q^2);~(\tau=p,n)$ are the structure functions for the proton and neutron which are evaluated independently following Refs.~\cite{Vermaseren:2005qc, Moch:2004xu}, i.e., without using the Callan-Gross
relation.

 Moreover, in a nucleus, the virtual photon
may interact with the virtual
mesons leading to the modification of the nucleon structure functions due to the additional contribution of the mesons. 
In the numerical calculations, we have considered the contribution from $\pi$ and $\rho$ mesons.
To obtain the contributions of $\pi$ and $\rho$ mesons to the structure functions we follow the similar procedure as in the 
case of nucleon with a difference that the spectral function is now replaced by the dressed meson propagator~\cite{Marco:1995vb, Haider:2016zrk, Haider:2015vea}. We find that

\begin{eqnarray} 
\label{pion_f21}
F_{{2 A,\pi(\rho)}}(x,Q^2)  &=&  -6\times a \int \, d^3 r \, \int \frac{d^4 p}{(2 \pi)^4} \, 
        \theta (p_0) ~\delta I m D_{\pi(\rho)} (p) \;2m_{\pi(\rho)}~\left(\frac{m_{\pi(\rho)}}{p_0~-~p_z~\gamma}\right)\times \nonumber \\
&&\left[\frac{Q^2}{(q_z)^2}\left( \frac{|{\bf p}|^2~-~(p_{z})^2}{2m_{\pi(\rho)}^2}\right)  
+  \frac{(p_0~-~p_z~\gamma)^2}{m_{\pi(\rho)}^2} \left(\frac{p_z~Q^2}{(p_0~-~p_z~\gamma) q_0 q_z}~+~1\right)^2\right] F_{{2,\pi(\rho)}}(x_{\pi(\rho)},Q^2),\\
\label{meson_f1}
 F_{{1A,\pi(\rho)}}(x,Q^2)  &=& -6\times a \times A M_N \int \, d^3 r \, \int \frac{d^4 p}{(2 \pi)^4} \, 
        \theta (p_0) ~\delta I m D_{\pi(\rho)} (p) \;2m_{\pi(\rho)}~\nonumber\\
     &\times&  \left[\frac{F_{1,\pi(\rho)} (x_{\pi(\rho)},Q^2)}{m_{\pi(\rho)}} +\frac{|{\bf p}|^2-p_z^2}{2(p_0q_0-p_zq_z)}\frac{F_{2,\pi(\rho)}(x_{\pi(\rho)},Q^2)}{m_{\pi(\rho)}}\right],~~
\end{eqnarray}
where $x_{\pi(\rho)}=\frac{Q^2}{-2 p \cdot q}$, $m_{\pi(\rho)}$ is the mass of pi(rho) meson and the constant factor $a$ is 1 in the case of $\pi$ meson and 2 in the case of $\rho$ meson~\cite{Haider:2016zrk, Haider:2015vea}.
$D_{\pi(\rho)}(p)$ is the meson propagator which is given by
\begin{eqnarray}
 D_{\pi(\rho)}(p)=[p_0^2-{\bf p}^2-m_{\pi(\rho)}^2-\Pi_{\pi(\rho)}(p_0,{\bf p})]^{-1},
\end{eqnarray}
where $\Pi_{\pi(\rho)}$ is the meson self energy defined in terms of the form factor $F_{{\pi(\rho)NN}}(p)$ and irreducible self energy $\Pi_{\pi(\rho)}^\ast$ as
\begin{eqnarray}\label{fpinn}
  \Pi_{\pi(\rho)}&=&\frac{\left(\frac{f^2}{m_\pi^2}\right)~c'_{\pi(\rho)}~F_{{{\pi(\rho)}NN}}^2(p) {\bf p}^2 \Pi_{\pi(\rho)}^\ast}{1-{f^2 \over m_\pi^2 }V_j^\prime \Pi_{\pi(\rho)}^\ast}\;,\;\mbox{where}\;\;
  F_{{{\pi(\rho)}NN}}(p) = \left(\frac{\Lambda^2 - m_{\pi(\rho)}^2}{\Lambda^2 + {\bf p}^2}\right).
\end{eqnarray}
In the above expression, $V_j^\prime=V_L^\prime$($V_T^\prime$) for the pi(rho) meson, 
are the longitudinal(transverse) part of spin-isospin interaction, respectively, the
expressions for which are taken from the Ref.~\cite{Marco:1995vb} with
$c'_{\pi}=1$ and $c'_{\rho}=3.94$, $\Lambda=1~GeV$ and $f=1.01$. 
 These parameters have been fixed in our earlier works~\cite{Haider:2016zrk, Haider:2015vea, Haider:2014iia, Haider:2011qs, Haider:2015nf, Haider:2015ic, SajjadAthar:2009cr}
while describing nuclear medium effects in the electromagnetic nuclear structure function $F_{2A} (x,Q^2)$ to explain the 
latest data from the JLab and other experiments performed using charged lepton scattering from several nuclear targets in the DIS region.

  For the pions, we have taken the pionic parton distribution functions given by 
 Gluck et al.~\cite{Gluck:1991ey} and for the rho mesons used the same PDFs as for the pions. In literature, there exists PDF parameterizations also for the mesons like that of
 Wijesooriya et al.~\cite{Wijesooriya:2005ir}, Sutton et al.~\cite{Sutton:1991ay}, Martin et al.~\cite{Martin:1998sq}, Conway et al.~\cite{Conway:1989fs}, etc. To see the dependence of mesonic structure functions on the 
  different PDFs parameterizations, we have also obtained the results by using the pionic PDFs parameterization given by Wijesooriya et al.~\cite{Wijesooriya:2005ir}.
We now define the total EM nuclear structure functions $F_{iA} (x,Q^2)$(i=1,2) which include the nuclear effects with spectral function and mesonic contributions as:
\begin{equation}\label{ftotal}
 F_{iA} (x,Q^2) =  F_{iA,N} (x,Q^2) +  F_{iA,\pi} (x,Q^2) + F_{iA,\rho} (x,Q^2)\;;\;\; i=1,2.
\end{equation}
and define $F_{LA} (x,Q^2)$ and $R_{A} (x,Q^2)$ 
in nuclear targets in analogy with $F_{LN} (x,Q^2)$ and $R_N(x,Q^2)$ as:
\begin{equation}\label{fla}
 F_{LA} (x,Q^2) = \left( 1+ {4 M_N^2 x^2 \over Q^2}\right)F_{2A} (x,Q^2) - 2 x F_{1A} (x,Q^2),
\end{equation}

\begin{equation}\label{rlar2a}
 R_{A} (x,Q^2)={F_{2A} (x,Q^2) \over 2 x F_{1A} (x,Q^2)}.
\end{equation}
\section{Results}\label{sec_results}

\begin{figure}
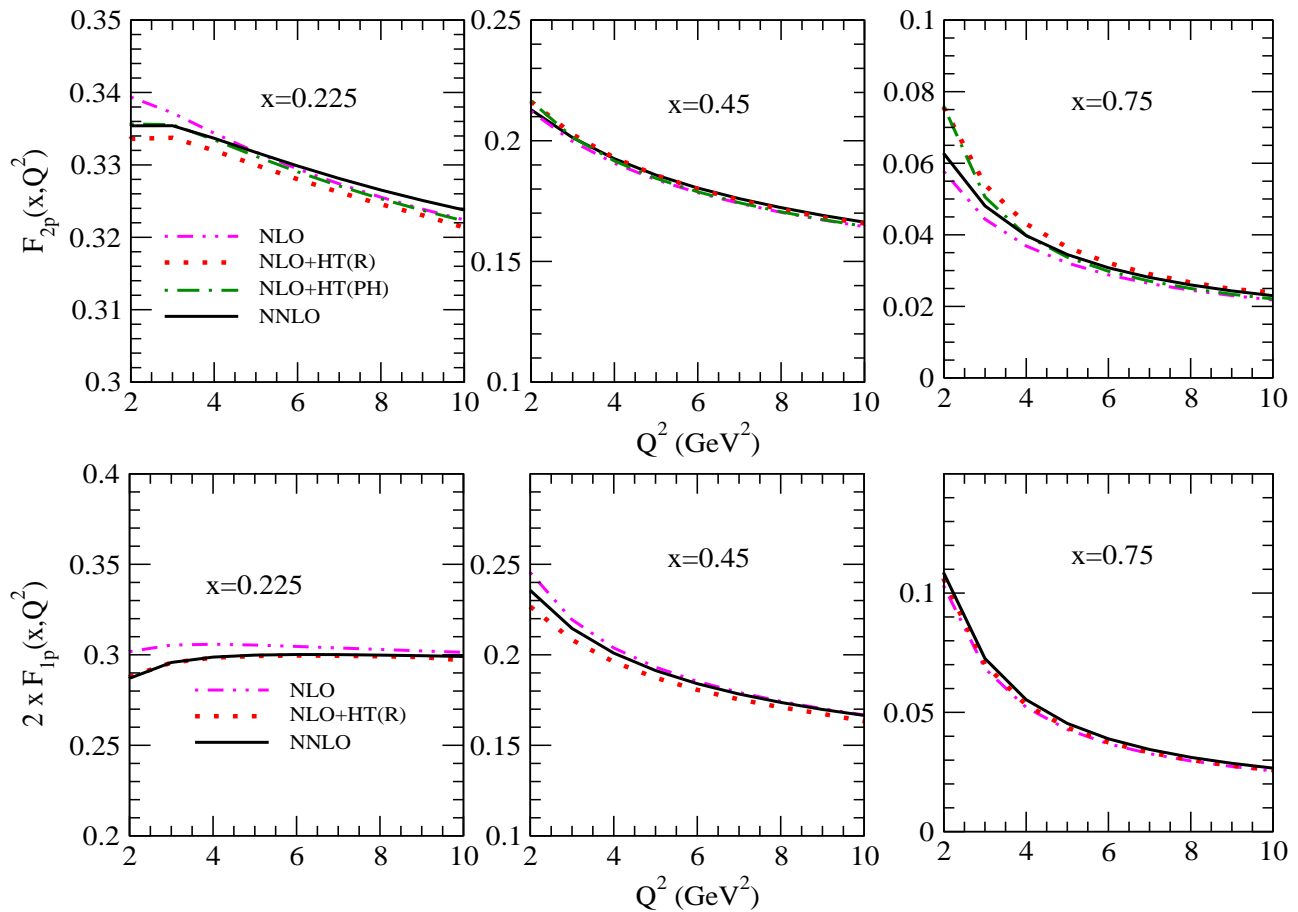

\begin{center}
 \includegraphics[height= 6 cm , width= 0.95\textwidth]{./fig/fig0_nuc_em_nodata.eps}\\
 \includegraphics[height= 6 cm , width= 0.95\textwidth]{./fig/fig0_nuc_em_nodata_f1_new.eps}
\end{center}
\caption{The results of $ F_{2p} (x,Q^2)$ and $ 2 x F_{1p} (x,Q^2)$ vs $Q^2$ are shown at different $x$ for the case of free proton. 
 The results are obtained at NLO with TMC(dashed-double dotted line) and also including HT effect following renormalon approach(dotted line) and phenomenological
 parameterization (dashed-dotted line). The results are also obtained at NNLO(solid line). }
\label{f1nf2n}
\end{figure}

\begin{figure}
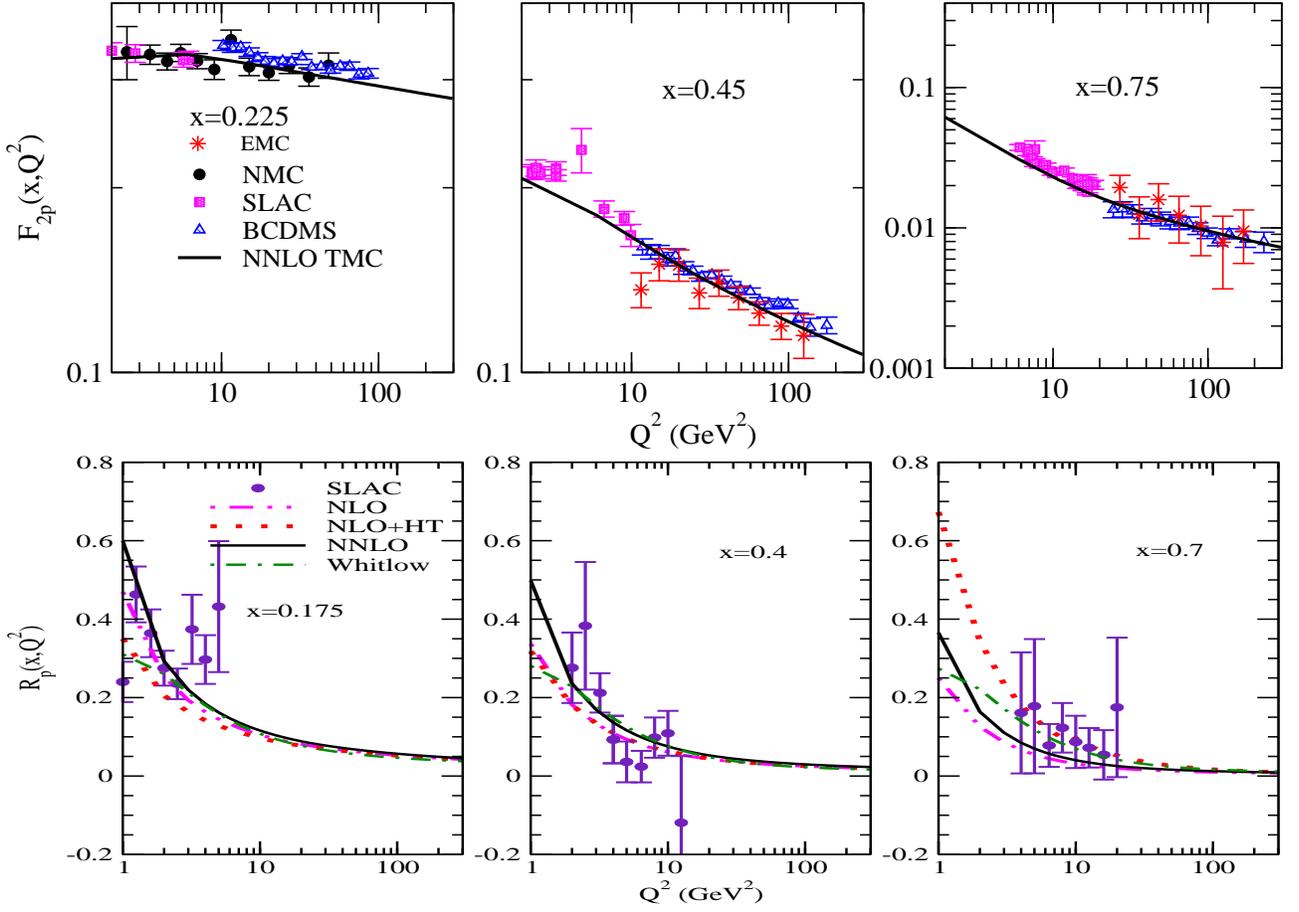

\begin{center}
 \includegraphics[height= 6 cm , width= 0.95\textwidth]{./fig/fig0_nuc_em.eps}\\
 \includegraphics[height= 6 cm , width= 0.95\textwidth]{./fig/rl_free.eps}\\
\end{center}
\caption{On the top panel the results of $ F_{2p} (x,Q^2)$ vs $Q^2$ are shown at different $x$ for the case of free proton. 
 On the bottom panel the results are presented for $R_p(x,Q^2)= \frac{F_{2p}(x,Q^2) (1 + \frac{4M_p^2 x^2}{Q^2})}{2xF_{1p}(x,Q^2)} - 1$.
 The results are obtained at NNLO(solid line), NLO(dashed-double dotted line) and also including HT effect following renormalon approach(dotted line).
 The dashed dotted line represents the results of the phenomenological fit of Whitlow et al.~\cite{Whitlow:1991uw}.
  The results are compared with the available experimental data from SLAC~\cite{Whitlow:1991uw}, BCDMS~\cite{Benvenuti:1989rh}, NMC~\cite{Arneodo:1996rv} and EMC~\cite{Aubert:1985fx} experiments. }
\label{prc:fig0}
\end{figure}
\begin{figure}
\begin{center}
 \includegraphics[height= 6 cm , width= 0.95\textwidth]{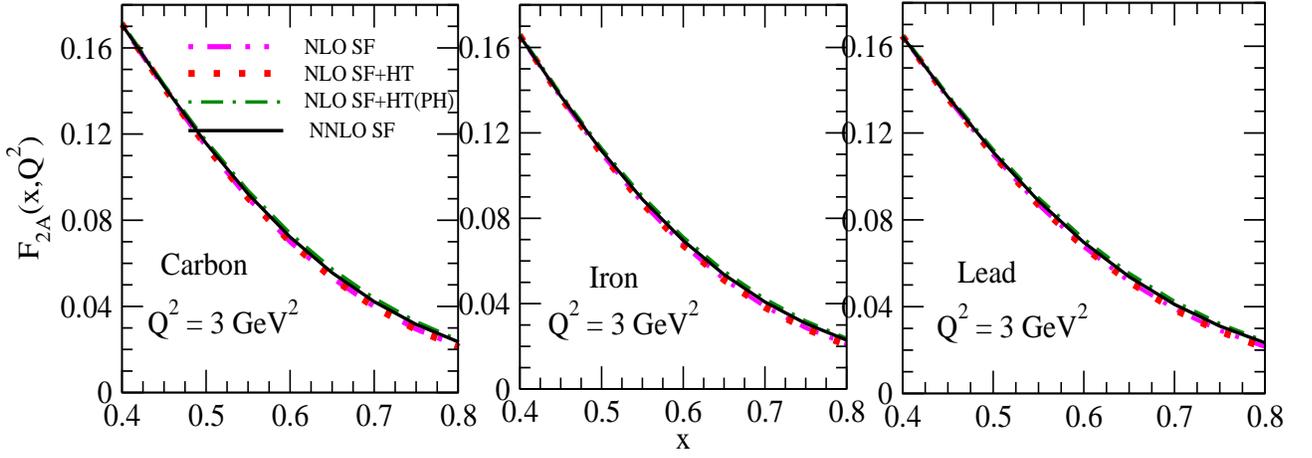}
\end{center}
 \caption{$F_{2A} (x,Q^2)$ ($A = ^{12}C$, $^{56}Fe$ and $^{208}Pb$) vs $x$ are shown at $Q^2=3~GeV^2$. 
  The results are obtained for the spectral function only without (dashed-double dotted line) and with the higher twist effect (renormalon approach: dotted line), using MMHT PDFs at NLO. The results are also obtained at NNLO using spectral function only (solid line). Dashed-dotted line is the result for the spectral function only obtained using the 
  phenomenological parameterization~\cite{Virchaux:1991jc} of HT effect at NLO. }
\label{prc:fig1}
\end{figure}

\begin{figure}
\begin{center}
 \includegraphics[height= 8 cm , width= 0.95\textwidth]{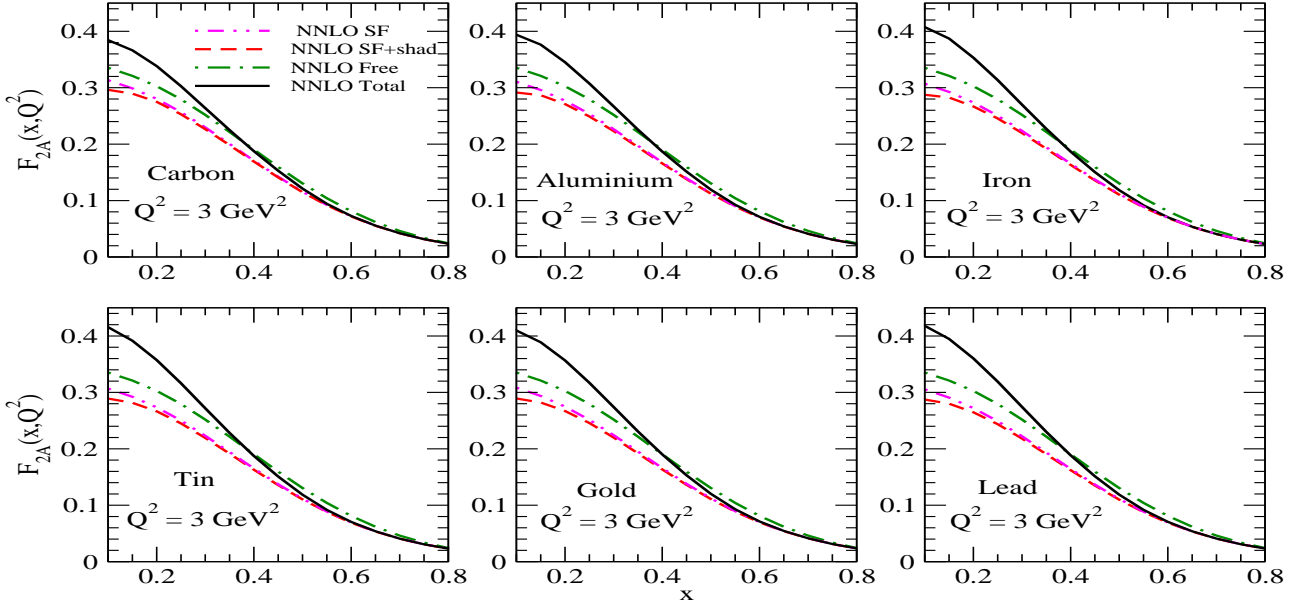}
\end{center}
 \caption{$F_{2A} (x,Q^2)$ ($A = ^{12}C$, $^{27}Al$, $^{56}Fe$, $^{118}Sn$, $^{197}Au$ and $^{208}Pb$) vs $x$ are shown at $Q^2=3~GeV^2$. 
 The results are obtained at NNLO using spectral function only (dashed-double dotted line), spectral function with shadowing effect (dashed line)
  and with the full model (solid line). The dashed-dotted line represents the results for the free nucleon case at NNLO.}
\label{prc:fig1p}
\end{figure}


 In Fig.~\ref{f1nf2n}, we present the numerical results for the proton structure functions $F_{2p}(x,Q^2)$ and $2xF_{1p}(x,Q^2)$ vs $Q^2$ at 
 the different values of $x$, for $Q^2 < 10~GeV^2$. The free nucleon structure functions $F_{iN} (x,Q^2)$ ($i=1,2,L$) at LO is obtained using the nucleon PDFs of 
 MMHT 2014~\cite{Harland-Lang:2014zoa}. 
 For the evolution of PDFs at NLO and NNLO from LO, we have followed the works of Vermaseren et al.~\cite{Vermaseren:2005qc}
 and Moch et al.~\cite{Moch:2004xu}. Then we have applied dynamical higher twist corrections following renormalon approach~\cite{Dasgupta:1996hh, Stein:1998wr} as well as phenomenological approach~\cite{Virchaux:1991jc} at the NLO. 
 All the theoretical results presented here are with the TMC effect~\cite{Schienbein:2007gr} which is found to be more pronounced in the region 
 of large $x$ and moderate $Q^2$.
  The numerical results are presented with (i) NLO, 
  (ii) NLO+HT(renormalon approach)~\cite{Dasgupta:1996hh, Stein:1998wr}, 
  (iii) NLO+HT(phenomenological approach)~\cite{Virchaux:1991jc}, and (iv) NNLO.
   It may be observed that in the case of $F_{2p}(x,Q^2)$ (top panel), the difference due to the HT effect (renormalon approach) from the results obtained without it is small at low $x$, however this difference becomes significant with the increase in $x$. For example,
   it is $\approx 2\%$ at $x=0.225$ and becomes $30\%$ at $x=0.75$ for $Q^2=2~GeV^2$ while this difference decreases to $<1\%$ at $x=0.225$ and $10\%$ at $x=0.75$ for $Q^2=6~GeV^2$. 
   The results at NLO with HT following renormalon approach
 are very close to the results obtained at NNLO except at high $x$ ($x>0.7$). For example, for $Q^2=2~GeV^2$ at $x=0.4$ the difference between the results with HT effect and the results at NNLO 
  is $\approx 2\%$ and it becomes $16\%$ at $x=0.75$. However, for $Q^2=6~GeV^2$ this difference reduces to $<1\%$ at $x=0.45$ and $4\%$ at $x=0.75$, respectively. 
  Furthermore, the results obtained with HT effect following the renormalon approach~\cite{Dasgupta:1996hh, Stein:1998wr} are in agreement within a percent ($<1\%$)
  with the results obtained by using the phenomenological prescription~\cite{Virchaux:1991jc} in the region of low and mid $x$.
  However, at high $x$ for example, at $x=0.75$ and for $Q^2=3~GeV^2$, there is a difference of about $6\%$ which gradually decreases with the increase in $Q^2$. We have also shown the results for $2 x F_{1p}(x,Q^2)$ vs $Q^2$ (bottom panel) for the same kinematical region as described above without and with the HT effect at NLO as well as compared them with the results obtained at NNLO. It is important to point out that the higher twist effect (renormalon approach) behaves differently for the free nucleon structure functions $F_{1p}(x,Q^2)$ and $F_{2p}(x,Q^2)$~\cite{Dasgupta:1996hh}. 
  From the Fig.~\ref{f1nf2n} (bottom panel), it may be observed that the results obtained without the HT effect differ from the results with HT effect at low $x$ and low $Q^2$, like there is a difference of $\approx 5\%$ at $x=0.225$ which reduces to $~\approx 3\%$ at $x=0.75$ for $Q^2=2~GeV^2$.
  Furthermore, we have observed that the results with HT effect obtained using the renormalon approach are in good agreement with the results at NNLO. For example,
  at $x=0.225$ this difference is $<1\%$ for $Q^2=2~GeV^2$ and becomes $2\%$ at $x=0.75$. Moreover, the effect of higher twist corrections becomes small with the increase in $Q^2$. 
  This is expected because higher twist effect has inverse power of $Q^2$, so at high $Q^2$ they should be less relevant.
  
 In Fig.~\ref{prc:fig0} (top panel), we present the numerical results for the proton structure function $F_{2p}(x,Q^2)$ vs $Q^2$ obtained using NNLO PDFs,
 at the different values of $x$ for a wide range of $Q^2$ and compared them with 
 the experimental data from SLAC~\cite{Whitlow:1991uw}, BCDMS~\cite{Benvenuti:1989rh}, NMC~\cite{Arneodo:1996rv} and EMC~\cite{Aubert:1985fx} experiments.
 We find reasonably good agreement of the theoretical results with the experimental data. 
 In this figure (bottom panel), we have also presented the results for $R_p(x,Q^2)=\frac{F_{2p}(x,Q^2)}{2 x F_{1p}(x,Q^2)}\left(1+\frac{4 M_p^2 x^2}{Q^2} \right) -1$ vs $Q^2$ at fixed values of $x$.
 These results are compared with the experimental data of SLAC~\cite{Whitlow:1991uw} as well as with the results obtained using the phenomenological 
 parameterization of Whitlow et al.~\cite{Whitlow:1991uw} and they are found to be consistent.

  We have calculated the nuclear structure functions $F_{1A}(x,Q^2),~F_{2A}(x,Q^2),~F_{LA}(x,Q^2)$ and the 
  ratio $R_{A}(x,Q^2)=\frac{F_{2A}(x,Q^2)}{2xF_{1A}(x,Q^2)}$ for 
 several nuclei like $^{12}C$, $^{27}Al$, $^{56}Fe$, $^{64}Cu$, $^{118}Sn$, $^{197}Au$ and $^{208}Pb$ 
 by using the nucleon spectral function in the nuclear medium taking into account medium
 effects like Fermi motion, Pauli blocking and nucleon correlations. The expressions for the 
 nuclear structure functions $F_{{2A,N}} (x_A,Q^2)$ and $F_{{1A,N}} (x_A, Q^2)$ with spectral function 
 given in Eqs.\ref{em_f2_noniso} and \ref{conv_WA1} are used for the numerical calculations,
 which we have called results with the spectral function(SF).
 The effect of the pion and rho mesons contributions i.e. $F_{{2 A,\pi(\rho)}}(x,Q^2)$(Eq.\ref{pion_f21}) and 
 $F_{{1 A,\pi(\rho)}}(x,Q^2)$(Eq.\ref{meson_f1}) are included using the pionic PDFs by Gluck et al.~\cite{Gluck:1991ey},
 and the effects of shadowing and the antishadowing following the works of Kulagin and Petti~\cite{Kulagin:2004ie}.
  This is the full nuclear model(Total) we are using, for which the numerical results are presented.
 
   In Fig.~\ref{prc:fig1}, we have presented the results for $F_{2A}(x,Q^2)$ vs $x~(0.4\le x \le 0.8)$, at a fixed value of $Q^2$ ($=3~GeV^2$) for 
 nuclear targets like $^{12}C,~^{56}Fe$ and $^{208}Pb$. These results are obtained 
 using the spectral function of the nucleons and the parton distribution functions at NLO, without (NLO SF) and with the higher twist effect (NLO SF+HT)
 following renormalon approach~\cite{Dasgupta:1996hh, Stein:1998wr} as well as 
 the phenomenological
 method (NLO SF+HT(PH)) of Virchaux et al.~\cite{Virchaux:1991jc}. These results are also obtained at NNLO using the spectral function only (NNLO SF). We find that the difference between the results obtained without and with the HT effect (renormalon approach) is $<1\%$ for low and mid region of $x$, however, for $x=0.8$ it is approximately $2\%$ in carbon and lead. Hence, it can be concluded that higher twist effect gets suppressed in the nuclear medium. Furthermore, the results of nuclear structure function $F_{2A}(x,Q^2)$ obtained at NNLO are also found to be 
 in good agreement with the results obtained at NLO with the HT effect.

 In Fig.~\ref{prc:fig1p}, the results for $F_{2A}(x,Q^2)$ vs $x$ are shown at $Q^2=3~GeV^2$ for the  different nuclei like
 $^{12}C$, $^{27}Al$, $^{56}Fe$, $^{118}Sn$, $^{197}Au$ and $^{208}Pb$ and are compared with the free nucleon structure function at NNLO.
 To explicitly show the effect of nuclear medium, the numerical results are obtained by using the spectral function only, including shadowing effect with the 
 spectral function, and with the full model. It is found that there is significant reduction in the nucleon structure function due to the nuclear medium effects as compared to the free nucleon case. For example, this reduction is $7\%$ in carbon at $x=0.2$, $10\%$ at $x=0.4$ and at $x=0.7$ it becomes $8\%$. We find that this reduction gets enhanced with the increase in the nuclear mass number, for example, in lead the reduction becomes $10\%$ at $x=0.2$, $14\%$ at $x=0.4$ and $11\%$ at $x=0.7$. Furthermore, 
 we find that the shadowing effect is very small in the kinematic region of our interest ($x \ge 0.1$), however, it is significant for $x<0.1$. For example,
  at $x=0.05$ (not shown here) the reduction due to the shadowing effect from the results with spectral function only is found to be $7\%$ in carbon, $\approx 13\%$ in iron and $15\%$ in lead. It implies that shadowing effect becomes prominent with the increase in the mass number. However, with the increase in $x$ it becomes small, for example, at $x=0.1$ it reduces to $5\%$ in carbon and $6\%$ in lead. When the mesonic contributions in our model are included with the spectral function the structure function gets increased at low and intermediate $x$ while for $x>0.6$ mesonic contributions  
 become small. For example, in carbon at $x=0.2$ the enhancement in the nuclear structure function due to the mesonic contribution is $\approx 20\%$ and it
 becomes $5\%$ at $x=0.5$. 
 Furthermore, we have also observed that mesonic contributions are nuclear mass dependent, e.g., in $^{56}Fe (^{208}Pb)$ the enhancement due to the mesonic contributions become
 $32\%(36\%)$ at $x=0.2$ and $7\%(8\%)$ at $x=0.5$. These medium effects are also found to be $Q^2$ dependent, for example, in carbon at $Q^2=6~GeV^2$ (not shown here), the enhancement due to the mesonic contributions are found to be small, like $16\%$ at $x=0.2$ and $3\%$ at $x=0.5$ respectively. 
 Hence, it can be concluded that nuclear medium effects depend on $x$, $Q^2$ and the mass of nuclear target. 

\begin{figure}
\begin{center}
 \includegraphics[height= 8 cm , width= 0.95\textwidth]{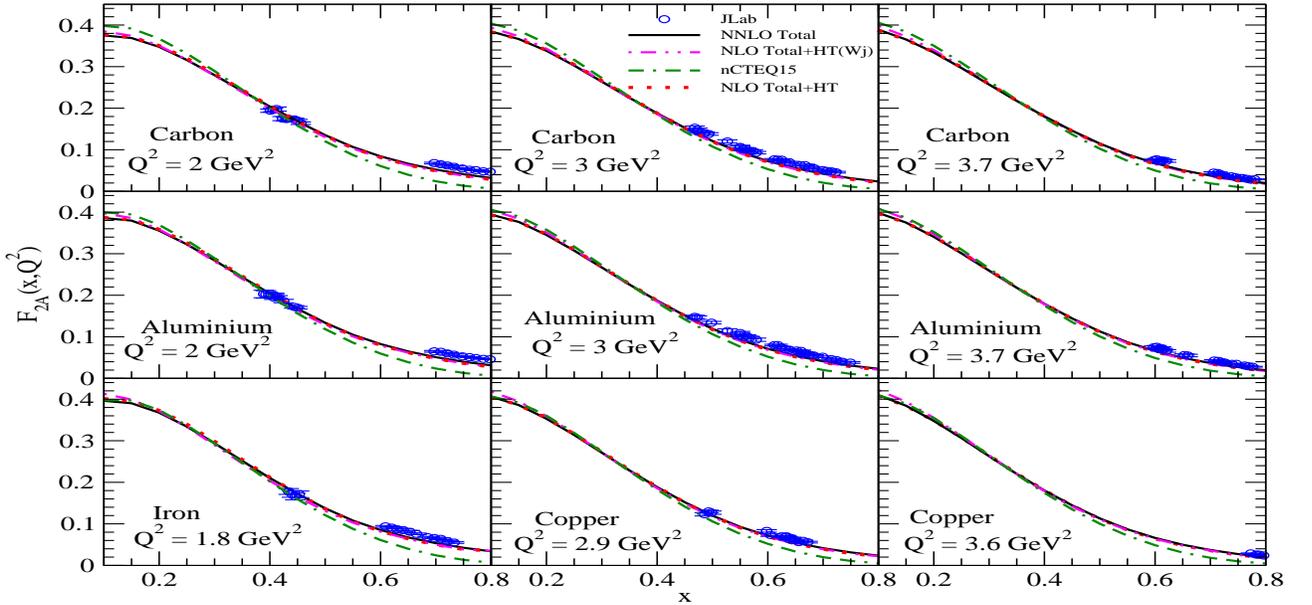}
\end{center}
\caption{$ F_{2A} (x,Q^2)$ ($A = ^{12}C$, $^{27}Al$, $^{56}Fe$ and $^{64}Cu$) vs $x$ are shown at different $Q^2$. 
 The results are obtained for the full model with HT effect (renormalon approach) using MMHT nucleonic PDFs and pionic PDFs of {\bf (i)} Gluck et al.~\cite{Gluck:1991ey} (dotted line), {\bf (ii)} Wijesooriya et al.~\cite{Wijesooriya:2005ir} (dashed-double dotted line) at NLO. Solid line is the results obtained at NNLO by using the MMHT nucleonic PDFs for the full model and double dashed-dotted line is the result obtained by using the 
  nCTEQ nuclear PDFs parameterization~\cite{Kovarik:2015cma}. The results are compared with the experimental data of JLab~\cite{Mamyan:2015th} (empty circles).}
\label{prc:fig2}
\end{figure}

\begin{figure}
\begin{center}
\includegraphics[height= 8 cm , width= 0.95\textwidth]{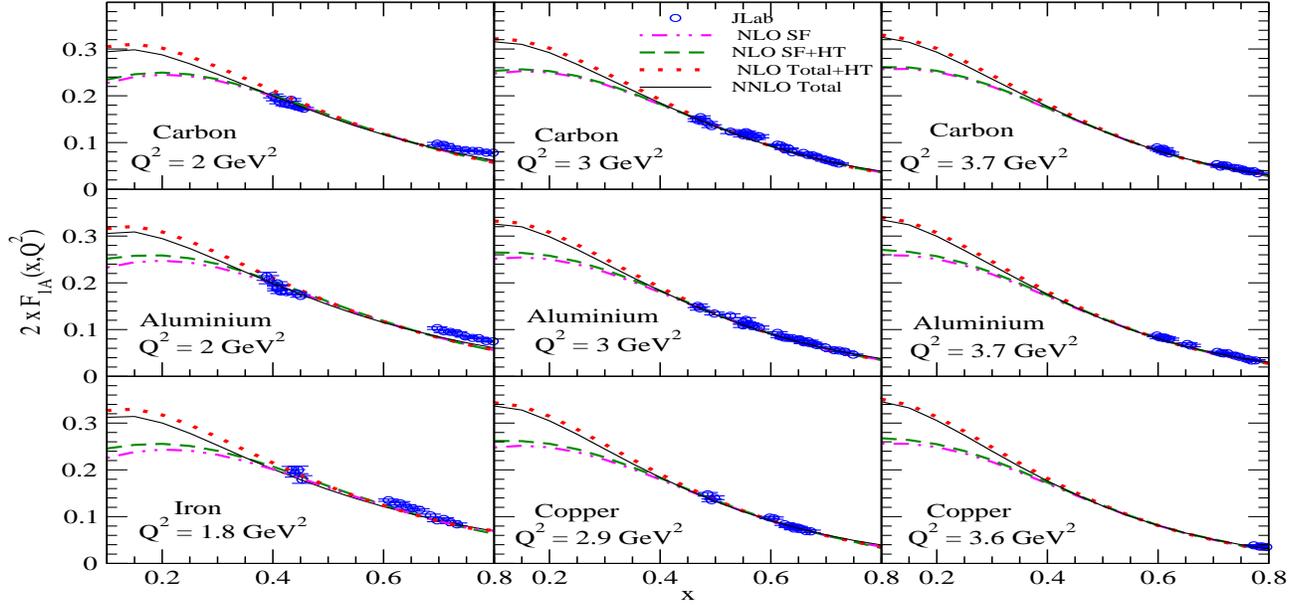}
\end{center}
\caption{$2 x F_{1A} (x,Q^2)$ ($A = ^{12}C$, $^{27}Al$, $^{56}Fe$ and $^{64}Cu$) vs $x$ are shown at different $Q^2$. 
 The results are obtained for the spectral function only without HT effect (dashed-double dotted line) and with HT effect (renormalon approach)
 for spectral function only (dashed line)
  and the full model (dotted line) using MMHT PDFs at NLO. Numerical results obtained by using the full model are also shown at NNLO (solid line) and are compared with the experimental data of JLab~\cite{Mamyan:2015th} (empty circles).}
\label{prc:fig3}
\end{figure}
\begin{figure}[t]
\begin{center}
\includegraphics[height= 8 cm , width= 0.95\textwidth]{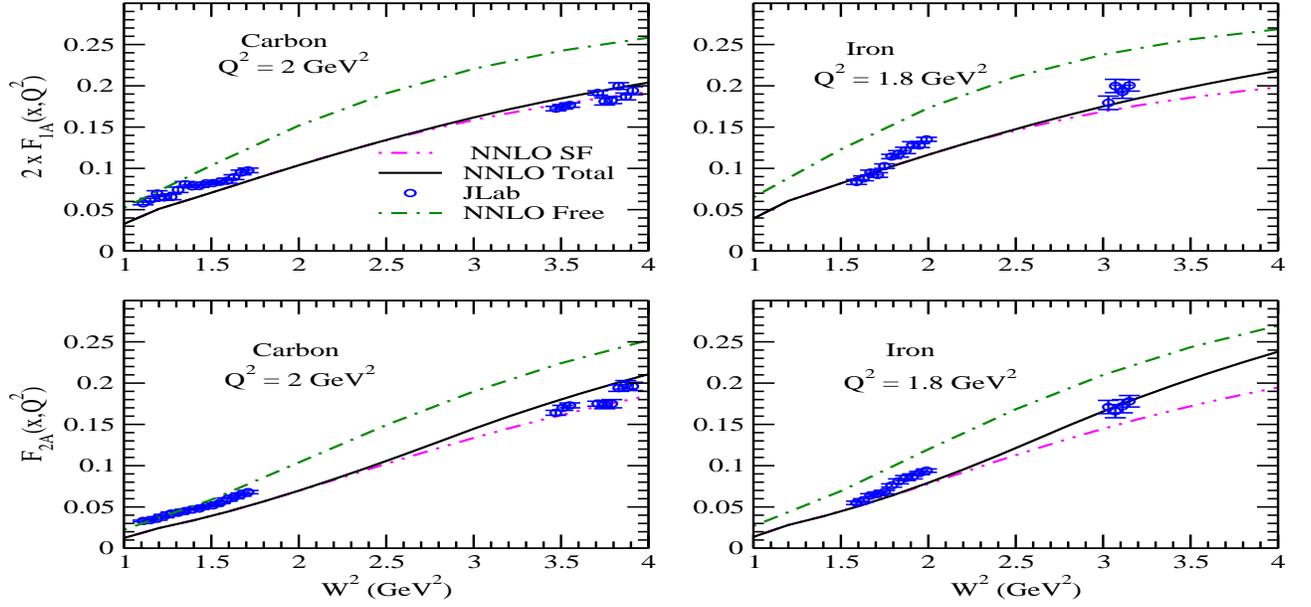}
\end{center}
\caption{$2 x F_{1A} (x,Q^2)$ and $F_{2A} (x,Q^2)$($A = ^{12}C$ and $^{56}Fe$) vs $W^2$ are shown at different $Q^2$. 
 The results are obtained for the spectral function only (dashed-double dotted line) 
  and for the full model (solid line) using MMHT PDFs at NNLO. The results are compared with the experimental data of JLab~\cite{Mamyan:2015th} (empty circles). Here we have also presented the results for the free nucleon case (dashed-dotted line).}
\label{prc:fig10}
\end{figure}
 
 In Fig.~\ref{prc:fig2}, we compare the results for $F_{2A}(x,Q^2)$ vs $x$ at different $Q^2$ ($\approx 2-4~GeV^2$) with the experimental observations of JLab~\cite{Mamyan:2015th}, for several nuclear targets like $^{12}C$, 
 $^{27}Al$, $^{56}Fe$ and $^{64}Cu$. Our theoretical results are 
 presented for the full model at NNLO, and at NLO with HT effect (renormalon approach). These results are compared
 with the phenomenological results given by nCTEQ group~\cite{Kovarik:2015cma} who have obtained nuclear PDFs for each nucleus separately. We find that our numerical 
 results with full model are reasonably in good agreement with the 
 nCTEQ results. To observe the dependence of pionic structure functions used in Eq.~(\ref{ftotal}), on the different pionic PDFs parameterizations we have also used the parameterization of Wijesooriya et al.~\cite{Wijesooriya:2005ir}. We have observed that the difference in the mesonic structure functions due to the parameterization of Wijesooriya et al.~\cite{Wijesooriya:2005ir} from the results
 obtained by using that of Gluck et al.~\cite{Gluck:1991ey} is within $1-3\%$ for all the nuclei under consideration. 
 Our theoretical results show a good agreement with the JLab experimental data~\cite{Mamyan:2015th} in the 
 region of intermediate $x$, however, for $x>0.6$ and $Q^2 \approx 2~GeV^2$ they slightly underestimate the experimental results. 
 Since the region of high $x$ and low $Q^2$ is the transition region of nucleon resonances and DIS, therefore, our theoretical results differ from the experimental data. However, with the increase in $Q^2$, theoretical results show better agreement with the experimental observations of 
 JLab~\cite{Mamyan:2015th} in the entire range of $x$.
 \begin{figure}
\begin{center}
\includegraphics[height= 7 cm , width= 0.95\textwidth]{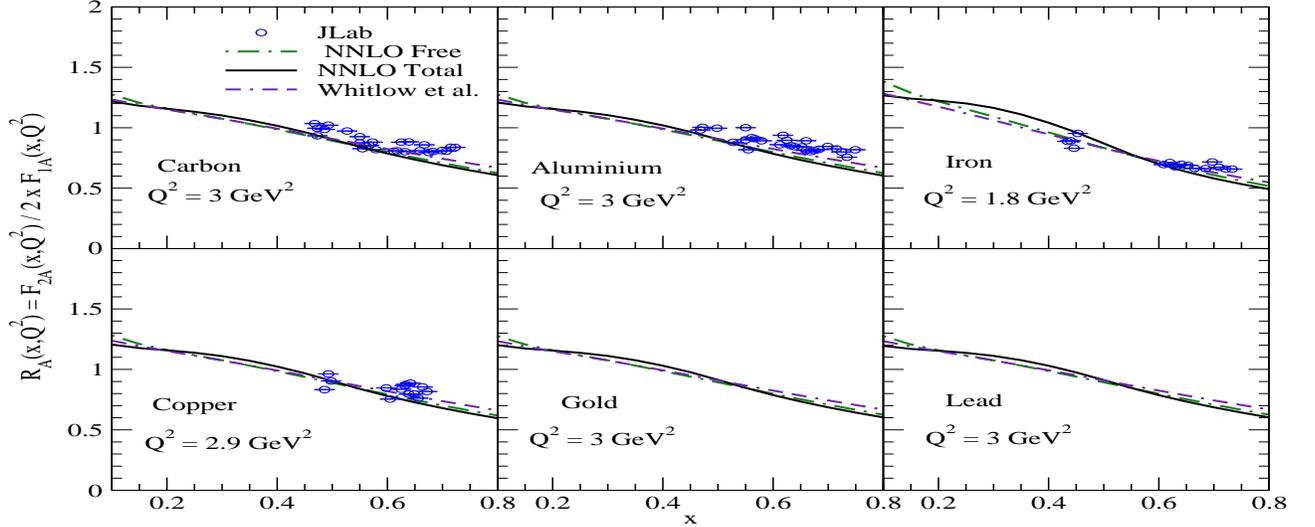}
\end{center}
\caption{Results for $R_{A} (x,Q^2)=\frac{F_{2A} (x,Q^2)}{2 x F_{1A} (x,Q^2)}$ ($A=$ $^{12}C$, $^{27}Al$, $^{56}Fe$, $^{64}Cu$, 
$^{197}Au$ and $^{208}Pb$) vs x are shown at different $Q^2$. Numerical 
results obtained using the full model(solid line) at NNLO, and are compared with the results of free nucleon case at NNLO (dashed-dotted line) and with the results obtained using the parameterization of Whitlow et al.~\cite{Whitlow:1991uw}(double dashed-dotted line). 
 These results are also compared with the available experimental data of the JLab~\cite{Mamyan:2015th}(empty circles). All the nuclear
 targets are treated as isoscalar.}
\label{prc:fig4}
\end{figure}

 \begin{figure}
\begin{center}
\includegraphics[height= 7 cm , width= 0.95\textwidth]{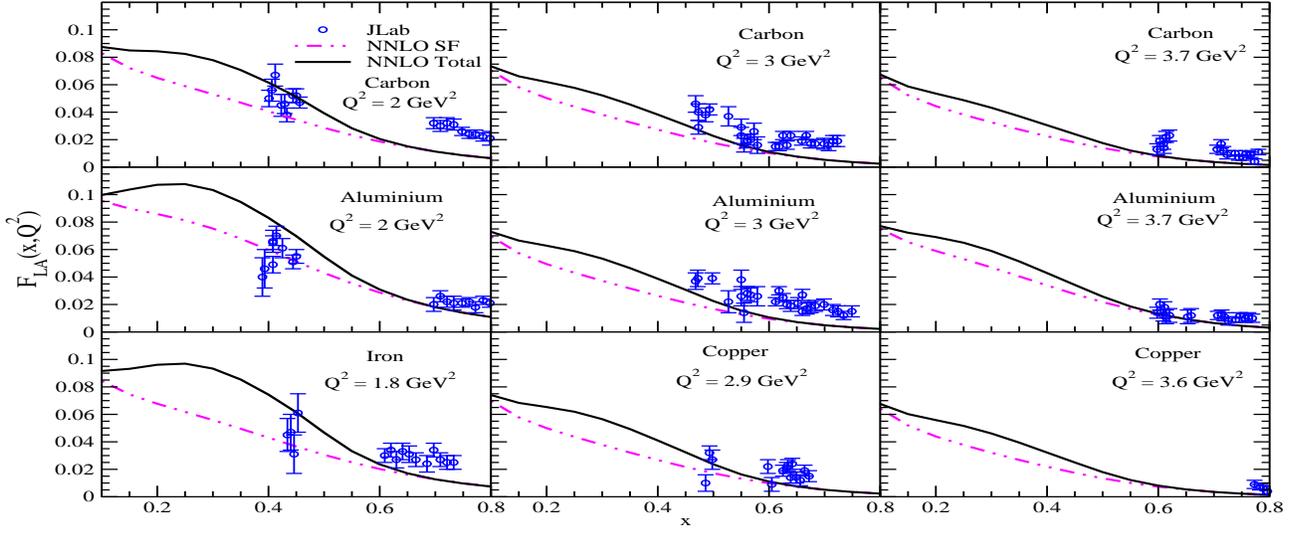}
\end{center}
\caption{The results are shown for the longitudinal structure function $F_{LA}(x,Q^2)$ ($A=^{12}C,~,^{27}Al,~^{56}Fe$ and $^{64}Cu$) vs $x$, for different $Q^2$ at NNLO with spectral function only (dashed-double dotted line) and with the full model (solid line). These results are compared with the experimental data (empty circles) of JLab~\cite{Mamyan:2015th}.}
\label{prc:fig6}
\end{figure}

\begin{figure}
\begin{center}
\includegraphics[height= 8cm , width= 0.95\textwidth]{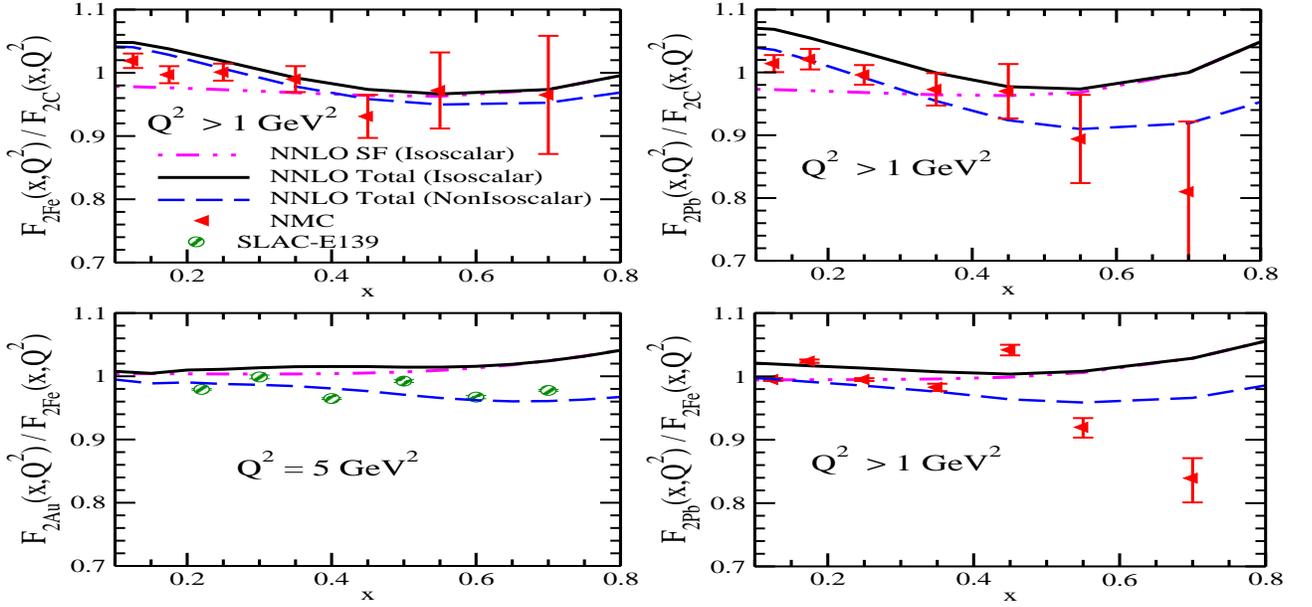}
\end{center}
\caption{The results are shown for the ratio of $F_{2A} \over {F_{2A'}}$ ($A=^{56}Fe,~^{197}Au,~^{208}Pb$ and $A'=^{12}C,~^{56}Fe$) vs x for $Q^2>1~GeV^2$ at NNLO for isoscalar and nonisoscalar nuclear targets. These results are compared with the  experimental data of SLAC~\cite{Gomez:1993ri} and NMC~\cite{Arneodo:1996rv}.}
\label{prc:fig9}
\end{figure}
 In Fig.~\ref{prc:fig3}, we present the results for $2 x F_{1A}(x,Q^2)$ vs $x$, for several nuclei in the intermediate mass range
 like $^{12}C,~^{27}Al,~^{56}Fe,$ and $^{64}Cu$. The results are presented at NLO for spectral function only without and with the
 higher twist effect (renormalon approach), as well as using the full model with HT effect (renormalon approach). We find that the effect of higher twist is more pronounced in the case of $2 x F_{1A}(x,Q^2)$
 than $F_{2A}(x,Q^2)$ structure function. 
 For example, in $^{64}Cu$ at $Q^2=2.9~GeV^2$ the difference in the results(without HT vs with HT) is $5\%$ at $x=0.1$ which decreases to $3\%$ at $x=0.2$. 
 At $Q^2=6~GeV^2$(not shown here), the difference in the results is $1\%$ at $x=0.1$ which becomes negligible at $x=0.2$.
 We also obtain the numerical results at NNLO using the full model which are found to be slightly different from the results obtained using the full model with HT effect 
 at NLO. The theoretical results are compared with the experimental data of JLab~\cite{Mamyan:2015th} and we find that numerical results underestimate the experimental data at high $x$ and low $Q^2$. However, for $0.7<x<0.8$ and $Q^2>2~GeV^2$ our results are in good agreement with the experimental observations.

In Fig.~\ref{prc:fig10}, we have presented the results for $2 x F_{1A}(x,Q^2)$ and $F_{2A}(x,Q^2)$ vs $W^2$,
  in $^{12}C$ at $Q^2= 2~GeV^2$ and in $^{56}Fe$ at $Q^2=1.8~GeV^2$ and compared the results with JLab data~\cite{Mamyan:2015th}. The theoretical results
are presented for the nuclear spectral function only as well as with the full model using MMHT nucleon PDFs at NNLO. We have also presented the results for the free
nucleon case. It may be observed that the present model with nuclear effects underestimates the experimental results at low $W$. It may be noticed from the figure that in the region of low $W^2<2.5~GeV^2$ which describes the resonance region dominated by some low lying resonances,  experimental data of JLab~\cite{Mamyan:2015th} overestimates our theoretical results. 
This may be due to the inadequacy of using DIS formalism at low $W$. In this region of low $W$ the contribution from the nucleon resonances like 
P$_{33}(1232)$, P$_{11}(1440)$, D$_{13}(1520)$, S$_{11}(1535)$, S$_{11}(1650)$, P$_{13}(1720)$,
etc. should better describe the experimental data.
However, for $W^2>2.5~GeV^2$, our numerical results which are obtained using the DIS formalism are found to be in reasonably good agreement. This behavior of nuclear structure functions supports our argument that for the region of low $Q^2<2~GeV^2$ and low $W \le 1.6~GeV$ a realistic calculation of nucleon resonances should be more appropriate as compared to the use of DIS formalism.


In Fig.~\ref{prc:fig4}, we have presented the results
for $R_{A} (x,Q^2)=\frac{F_{2A} (x,Q^2)}{2 x F_{1A} (x,Q^2)}$($A=$ $^{12}C$, $^{27}Al$, $^{56}Fe$, $^{64}Cu$, 
$^{197}Au$ and $^{208}Pb$) vs $x$ at different $Q^2$. Numerical 
results are obtained using the full model at NNLO, and are compared with the results for the free nucleon case at NNLO.
 Moreover, we have also presented the results of Whitlow et al.~\cite{Whitlow:1991uw}, who have parameterized the nucleon structure function $F_{1N}(x,Q^2)$ by using SLAC experimental
 data for $e^{-}-p$ and $e^{-}-d$ scattering processes. 
 These results are also compared with the available experimental data of the JLab~\cite{Mamyan:2015th} which are corrected for the isoscalar nuclear targets. The agreement with the experimental results as well as with the Whitlow's parameterization is satisfactory.

In Fig.~\ref{prc:fig6}, we have presented the results of longitudinal structure function $F_{LA}(x,Q^2)$ vs $x$, at different $Q^2$ for 
 several nuclear targets like $^{12}C,~^{27}Al,~^{56}Fe,$ and $^{64}Cu$. These results are presented for the spectral function only as well as with 
 the full model, using nucleon PDFs at NNLO. These results are compared with the experimental data of the JLab~\cite{Mamyan:2015th}. The agreement with the experimental result is reasonably good except at very low $Q^2<2~GeV^2$.

In Fig.~\ref{prc:fig9}, we have obtained $\frac{F_{2~^{56}Fe}(x,Q^2)}{F_{2~^{12}C}(x,Q^2)}$ and $\frac{F_{2~^{208}Pb}(x,Q^2)}{F_{2~^{12}C}(x,Q^2)}$,
$\frac{F_{2~^{197}Au}(x,Q^2)}{F_{2~^{56}Fe}(x,Q^2)}$ and $\frac{F_{2~^{208}Pb}(x,Q^2)}{F_{2~^{56}Fe}(x,Q^2)}$ using spectral function as well as the full model assuming the 
nuclear targets to be isoscalar. The results are also presented for the full model when $^{56}Fe,~^{197}Au$ and $^{208}Pb$ are treated as non-isoscalar 
nuclear targets where we normalize the spectral function to the proton and neutron numbers, separately. 
We obtain the ratio $\frac{F_{2~Fe}(x,Q^2)}{F_{2~C}(x,Q^2)}$ and $\frac{F_{2~Pb}(x,Q^2)}{F_{2~C}(x,Q^2)}$ for $1 \le Q^2 \le 66 ~GeV^2$, by first
 assuming $^{208}Pb$ and $^{56}Fe$ to be isoscalar targets, and then both of them as nonisoscalar targets, 
 and find the isoscalarity effect to be $<1\%(3\%)$ and $\approx 3\%(9\%)$ for $^{56}Fe(^{208}Pb)$ at $x=0.125$ and at $x=0.8$, respectively. We have also presented the ratio $\frac{F_{2~Pb}(x,Q^2)}{F_{2~Fe}(x,Q^2)}$ 
 assuming $^{208}Pb$ and $^{56}Fe$ to be isoscalar targets, as well as nonisoscalar targets. We find the
 isoscalarity effect to be $2\%$ at $x=0.125$ which increases to 
$\approx 7\%$ at $x=0.8$ for $1 \le Q^2 \le 66 ~GeV^2$. Similarly in the case of $\frac{F_{2~Au}(x,Q^2)}{F_{2~Fe}(x,Q^2)}$ at $Q^2=5~GeV^2$, the
 isoscalarity effect is found to be $1\%$ at $x=0.1$ which increases to 
$7\%$ at $x=0.8$. These results are also compared with the experimental data from SLAC~\cite{Gomez:1993ri} and NMC~\cite{Arneodo:1996rv} experiments and are found to be in fair agreement with them. 

\section{Summary and Conclusion}\label{summary}
 In this work, we have studied the effect of non-perturbative and higher order perturbative corrections on the evaluation of nucleon structure functions and its 
 implications in the calculations of the nuclear structure functions.
 For the nucleon structure functions which are described in terms of nucleon PDFs the evaluations are made at NLO with HT as well as at NNLO. The nuclear structure 
 functions are obtained using a microscopic nuclear model and the effects of 
 the Fermi motion, binding energy, nucleon correlations, mesonic contribution and shadowing are considered. We have also studied the impact of these corrections on the
 Callan-Gross relation in free nucleons and nuclei.
 
 We find that:
  \begin{enumerate}
  \item The nucleon structure functions $F_{2N}(x,Q^2)$ and $2 x F_{1N}(x,Q^2)$ get modified at high $x$ and low $Q^2$ due to the inclusion of higher twist effect when evaluated at NLO. However, for low $x$ region the impact of HT effect in $2 x F_{1N}(x,Q^2)$ is found to be more pronounced than in the case of $F_{2N}(x,Q^2)$. The HT effect decreases with the increase in $Q^2$.
\item The effect of higher twist in nuclei is small in $F_{2A}(x,Q^2)$ and the results obtained at the NNLO are very close to the NLO+HT results.
 Qualitatively the effect of HT on the $2 x F_{1A}(x,Q^2)$ evaluation is similar to what has been observed in $F_{2A}(x,Q^2)$, however, quantitatively
 the effect is not too small specially at low $x$ and low $Q^2$. This is the same finding as that observed in the case of nucleon structure functions.
 
 \item The inclusion of nuclear medium effects leads to a better description of the experimental data from JLab\cite{Mamyan:2015th}, 
 SLAC~\cite{Gomez:1993ri} and NMC~\cite{Arneodo:1996rv} in various nuclei in a wide range 
 of $x$ and $Q^2$. At high $Q^2$ the experimental results are well reproduced, while at low $Q^2$($\le~ 2 ~GeV^2$) we underestimate the experimental 
 data for $x \ge 0.6$, where resonance contribution may be important.
 
\item In nuclei there is very small deviation in the Callan Gross relation($R_A(x,Q^2)$)
 from the free nucleon value due to the nuclear medium effects at low and moderate $Q^2$.
 The present results are in the right direction 
 to give a better description  of the available experimental data but underestimates them for $x > 0.6$. 

 \item The use of 
 DIS formalism to calculate the contribution of $2 x F_{1A}(x,Q^2)$, $F_{2A}(x,Q^2)$, $R_A(x,Q^2)$
 in the region of low $W$ and low $Q^2$ underestimates the experimental 
 results. In this
 kinematic region an explicit calculation of $R_A(x,Q^2)$ including the contribution arising due to the resonance excitation of $\Delta(1232)$ and $N^*$(1440) in the nuclear medium 
 should be more appropriate. 
 \end{enumerate}
\section*{Acknowledgment}   
M. S. A. and S. K. S. are thankful to Department of Science and Technology (DST), Government of India for providing 
financial assistance under Grant No. EMR/2016/002285. I.R.S. acknowledges support from Spanish Ministerio de Economia y Competitivedad under grant No. FIS2017-85053-C2-1-P, and by Junta de Andalucia (Grant No. FQM-225).

\end{document}